\documentclass[
    twocolumn,
    amsfonts,
    ammsymb,
    amsmath,
    nobibnotes,
    aps,
    prd,
    nofootinbib,
    superscriptaddress
]{revtex4-2}

\usepackage{amsmath}
\usepackage{amsfonts}
\usepackage{amssymb}
\usepackage{booktabs}
\usepackage[T1]{fontenc}
\usepackage{graphicx}
\usepackage[colorlinks=true, linkcolor=blue, urlcolor=blue, citecolor=blue]{hyperref}
\usepackage{siunitx}
\usepackage{xcolor}
\usepackage{orcidlink}

\newcommand{\A}{\mathcal{A}}
\newcommand{\cost}{\mathcal{C}}
\newcommand{\depth}{\mathcal{D}}
\newcommand{\F}{\mathcal{F}}

\newcommand{\HGauss}{\mathcal{H}_{\mathrm{G}}}
\newcommand{\HSignal}{\mathcal{H}_{\mathrm{S}}(\lambda, \A)}
\newcommand{\muq}{\mu^{90\%}}
\newcommand{\N}{\mathcal{N}}

\newcommand{\Nseg}{N_{\mathrm{seg}}}
\newcommand{\Nwalk}{N_{\mathrm{w}}}
\newcommand{\Ntemp}{N_{\mathrm{t}}}
\newcommand{\Ntot}{N_{\mathrm{tot}}}
\newcommand{\pfa}{p_{\mathrm{fa}}}
\newcommand{\pfd}{p_{\mathrm{fd}}}
\newcommand{\pyfstat}{\texttt{pyfstat}}
\newcommand{\ptemcee}{\texttt{ptemcee}}
\newcommand{\Sn}{S_{\mathrm{n}}}
\newcommand{\Tcoh}{T_{\rm coh}}
\newcommand{\Tobs}{T_{\rm obs}}
\newcommand{\Tsft}{T_{\rm SFT}}
\newcommand{\tasc}{t_{\mathrm{asc}}}
\newcommand{\tmid}{t_{\mathrm{mid}}}
\newcommand{\thresh}{2\F_{\mathrm{thr}}}
\newcommand{\tref}{t_{\mathrm{ref}}}

\DeclareMathOperator*{\argmin}{arg\,min}

\usepackage{graphicx}	\usepackage{amsmath}	\usepackage{amssymb}	\usepackage{siunitx}
\usepackage{xcolor}

\graphicspath{{figures/}}

\newcommand{\UIB}{
Departament de F\'isica, Universitat de les Illes Balears, 
IAC3 -- IEEC, Carretera Valldemossa km 7.5, E-07122 Palma, Spain
}
\newcommand{\milano}{Dipartimento di Fisica ``G. Occhialini'', 
Universit\`a degli Studi di Milano-Bicocca, Piazza della Scienza 3, 20126 Milano, Italy}
\newcommand{\infn}{INFN, Sezione di Milano-Bicocca, 
Piazza della Scienza 3, 20126 Milano, Italy}

\begin{document}

\title{
    Toward a computationally-efficient follow-up pipeline for\\blind continuous gravitational-wave searches
}

\author{Lorenzo Mirasola\,\orcidlink{0009-0004-0174-1377}}
\email{lorenzo.mirasola@ca.infn.it}
\affiliation{Physics Department, Universit\`a degli Studi di Cagliari, Cagliari 09042, Italy}
\affiliation{INFN sezione di Cagliari, Cagliari 09042, Italy}
\author{Rodrigo Tenorio\,\orcidlink{0000-0002-3582-2587}}
\email{rodrigo.tenorio@unimib.it}
\affiliation{\milano}
\affiliation{\infn}
\affiliation{\UIB}

\begin{abstract}
    The sensitivity of continuous gravitational-wave (CW) searches for unknown neutron stars (NSs)
    is limited by their parameter space breadth. To fit within reasonable computing budgets,
    hierarchical schemes are used to identify interesting candidates using affordable methods.
    The resulting sensitivity depends on the number of candidates selected to follow-up.
    In this work, we present a novel framework to evaluate the effectiveness of stochastic CW
    follow-ups. Our results allow for a significant reduction of the computing cost
    of~\pyfstat{}, a well-established follow-up method. We also simplify the setup of multistage 
    follow-ups by removing the need for parameter-space metrics. The study was conducted on Gaussian
    and real O3 Advanced LIGO data covering both isolated and binary sources. These results will have
    a positive impact on the sensitivity of all-sky searches in the forthcoming observing runs of 
    the LIGO-Virgo-KAGRA collaboration.
\end{abstract}

\maketitle

\section{Introduction\label{sect:intro}}
Continuous gravitational waves (CWs) are long-lasting gravitational-wave signals
expected to be produced by rapidly spinning nonaxisymmetric neutron stars (NSs)~\cite{Riles:2022wwz}.
Although currently undetected, CWs will extend our knowledge on galactic NSs,
which so far have only been observed
electromagnetically~\cite{Reed_2021, 2005AJ....12ss9.1993M}.

All-sky searches attempt to detect CW signals from unknown sources. The resulting
parameter-space breadth makes fully coherent matched filtering of year-long datasets
computationally unfeasible~\cite{Wette:2023dom}; 
semicoherent methods, which allow for coarser parameter-space discretizations, are used instead.
Interesting candidates are then followed up using more sensitive methods~\cite{Tenorio:2021wmz}.
These hierarchical schemes make all-sky searches more robust to stochastic NS physics,
such as glitches~\cite{Ashton:2017wui} or accretion-induced spin wandering~\cite{Mukherjee:2017qme}.

A favorable strategy to increase the sensitivity of a search without imposing stricter constraints
on the signal model is increasing the number of candidates to follow up: generally,
the more candidates are selected, the weaker a signal can be before being dismissed.
As a result, all-sky search sensitivity is a monotonic function of the available follow-up computing budget.
It is therefore fundamental to develop computationally efficient follow-up strategies to achieve
a first CW detection from an unknown source.

In this work, we develop a novel framework to characterize the effectiveness of a generic
CW follow-up as a function of signal mismatch and computing cost. The results of this study 
allow us to justify a five to ten-fold computing-cost reduction of a generic follow-up stage 
using~\ptemcee{}~\cite{2016MNRAS.455.1919V,2013PASP..125..306F} as implemented 
in~\pyfstat{}~\cite{PhysRevD.97.103020, Keitel:2021xeq}, a state-of-the-art generic CW follow-up
package based on Markov Chain Monte Carlo (MCMC) methods. Additionally, we simplify the  setup
of multistage follow-up schemes by removing the need for a full parameter-space metric~\cite{Ashton:2017wui}. 

This framework is applied to simulated all-sky searches for unknown NSs, both isolated and in binary systems,
using simulated Gaussian noise and real data from the O3 LIGO-Virgo-KAGRA observing
run~\cite{Tse:2019wcy,Virgo:2019juy,2020PhRvD.102f2003B,LIGO:2021ppb,
Virgo:2022ysc,KAGRA:2023pio}. We demonstrate the impact of our developments on real data
by estimating the sensitivity improvement on an all-sky search in Advanced LIGO O3 data~\cite{PhysRevD.106.102008}
due to the selection of a higher number of candidates. 

The low-cost follow-up strategy discussed here is especially applicable to short-coherence CW searches,
such as those described in~\cite{Krishnan:2004sv, Astone:2014esa, Covas:2019jqa}, which operate by 
identifying narrow parameter-space regions of interest on which to deploy a more sensitive method.
In this sense, this work complements that of~\cite{Covas:2024pam}, where a different follow-up approach,
targeting broader parameter-space regions, is discussed.

The paper is structured as follows: in Sec.~\ref{sect:follow_up_introduction}, we revisit the basics of 
CW MCMC follow-ups and discuss a new approach to evaluate their sensitivity.
In Sec.~\ref{sect:evaluating_effectiveness}, we formulate a multiobjective optimization
problem to construct effective MCMC configurations. The effectiveness of different follow-up
configurations is evaluated in Sec.~\ref{sect:optimizing_setup} by means of
an injection campaign on Gaussian and real O3 data.
The sensitivity impact of these improvements on a real search is discussed in 
Sec.~\ref{sect:impact}. We conclude in Sec.~\ref{sect:conclusions}.
 
\section{Follow-up of continuous-wave candidates\label{sect:follow_up_introduction}}

Blind searches identify interesting CW candidates using a computationally
affordable method to then follow them up using a  more sensitive 
approach. The purpose of this hierarchical approach is to gradually reduce
the false-alarm probability associated with a CW candidate while maintaining
the parameter-space size under 
control~\cite{Brady:1998nj, Krishnan:2004sv,PhysRevD.72.042004,Prix:2012yu}.

Several approaches have been proposed to follow-up CW candidates using long coherence
times~\cite{Shaltev:2013kqa,PhysRevD.89.124030, PhysRevD.93.044058,PhysRevD.94.122006,
PhysRevD.100.024004, Steltner:2020hfd, Steltner:2023cfk, Astone:2014esa}.
These make use of lattice template banks, which are feasible for isolated
sources but are impractical for sources in binary systems~\cite{PhysRevLett.124.191102}.
Lately, however,  MCMC strategies have proven successful to follow-up
candidates produced by a broad class of CW searches~\cite{PhysRevD.97.103020,
PhysRevD.104.084012,Covas:2024pam}.

In this section, we revisit the fundamentals of stochastic multistage CW follow-up analyses
of narrow parameter-space regions. 
The semicoherent $\F$-statistic and the basics of single-stage MCMC
follow-ups are introduced in Secs.~\ref{subsec:prolegomena}~and~\ref{subsec:single-stage}. 
In Sec.~\ref{subsec:stopping} we discuss, for the first time, the sensitivity of a follow-up in
terms of its false-alarm and false-dismissal probabilities.
In Sec.~\ref{subsect:ladder} we propose a novel and simpler strategy to construct
multistage follow-up schemes avoiding the use of a full parameter-space metric.

\subsection{CW prolegomena\label{subsec:prolegomena}}

CW signals as emitted by rapidly spinning nonaxisymmetric NSs can be described using
two sets of parameters: the~\emph{amplitude}  parameters $\A$ and 
the~\emph{phase-evolution}  parameters $\lambda$. 
This split is motivated by the detector's  response to such signals\begin{equation}
    h(t;\lambda, \A) = \sum_{\mu=0}^{3} \A^\mu\,h_\mu(t;\lambda) \,,
    \label{eq:signal}
\end{equation}
where $t$ refers to the time at the GW detector or Solar System Barycenter (SSB),
depending on the specific pipeline implementation, and $\A^{\mu}=\A^{\mu}(\A)$ are 
time-independent functions of the amplitude parameters
that can be treated analytically~\cite{PhysRevD.58.063001,PhysRevD.72.063006,
Prix:2009tq,Whelan2013NewCF}.
In the case of a NS sustaining a quadrupolar deformation, the amplitude parameters include
the initial phase $\phi_0$, the polarization angle $\psi$, the cosine of the inclination angle
with respect to the line of sight $\cos\iota$ and the nominal amplitude $h_0$, which depends
on the specific CW emission mechanism~\cite{Wette:2023dom}.

For an isolated NS, the phase-evolution parameters $\lambda_{\mathrm{i}}$ 
include the GW frequency $f_0$, the linear spindown parameter $f_1$ at a 
fiducial reference time $\tref$ (further spindown terms can be included 
depending on the age of the source), and the sky position, which we parameterize
using the equatorial angles $\hat{n} = (\alpha, \delta)$.
The instantaneous frequency as measured by a detector is
\begin{equation}
    f(t; \lambda_{\mathrm{i}}) = 
    \left[ f_0 + f_1 (t-\tref) \right] \cdot 
    \left(1+\frac{\vec{v}(t)}{c} \cdot \hat{n} \right) \,,
\end{equation}
where $\vec{v}$ is the velocity of the detector with respect to the SSB.

For a NS in a circular binary system, phase-evolution parameters $\lambda_{\mathrm{b}}$
include the orbital period $P$ or orbital frequency $\Omega = 2 \pi / P$, the projected
semimajor axis (in light seconds) $a_{\mathrm{p}}$, and the time of passage through 
the ascending node $\tasc$~\cite{PhysRevD.91.102003}:
\begin{equation}
    f(t; \lambda_{\mathrm{b}}) = f_0 \cdot 
    \left(1
    + \frac{\vec{v}(t)}{c} \cdot \hat{n}
    - a_{\mathrm{p}} \Omega \cos\left[\Omega (t - \tasc)\right]
    \right)\,.
    \label{eq:f_bin}
\end{equation}
These searches tend to omit spindown terms and  eccentricity as their effect on a circular-orbit template is
unresolvable for typical search
stages~\cite{PhysRevLett.124.191102,PhysRevD.103.064017,Covas:2022rfg}.

Most all-sky CW searches operate by comparing the detector data $x$ to a bank of
waveform templates \mbox{$\{ \lambda_j, j=1,\dots,N_{\mathrm{T}}\}$} by means 
of a detection statistic. The long duration of these signals makes fully coherent 
matched filtering  unaffordable; instead, these searches make use of semicoherent methods,
which split the data into $\Nseg$ segments with a certain duration $\Tcoh$ from which
a detection statistic is computed (see~\cite{Wette:2023dom} for a recent review).

In a Bayesian context~\cite{Prix:2009tq,Prix:2011qv}, semicoherent detection statistics can be related 
to a targeted marginal Bayes factor $B_{\mathrm{SG}}(x;\lambda)$ comparing the 
Gaussian noise hypothesis $\HGauss$,  under which the data consist of Gaussian noise samples 
$x = n$, versus the signal hypothesis $\HSignal$, which includes a CW signal with a specific 
set  of parameters $x = n + h(\lambda, \A)$:
\begin{equation}
    B_{\mathrm{SG}}(x;\lambda) =
    \int \mathrm{d}\A \, \frac{p(x| \HSignal) }{p(x|\HGauss)}p(\A)\,.
    \label{eq:bayes_fact}
\end{equation}
(Note that we marginalize the likelihood \emph{ratio} rather than the signal-hypothesis
likelihood itself as $\HGauss$ is independent of $\A$). By splitting the dataset
into disjoint segments and independently marginalizing the amplitude parameters in each of them we obtain 
the semicoherent $\F$-statistic~\cite{PhysRevD.58.063001,PhysRevD.72.063006,Prix:2011qv}
\begin{equation}
    B_{\mathrm{SG}}(x;\lambda) \propto e^{\hat{\F}(\lambda; x)} \,,
\end{equation}
\begin{equation}
    \hat{\F}(\lambda; x)= \sum_{s = 1}^{\Nseg} \F_{s}(\lambda;x) \,,
\end{equation}
where $\F_{s}$ refers to the fully coherent $\F$-statistic computed on segment $s$.
Intuitively, splitting the data into segments trades ``height'' for ``breadth'', in the sense
that the $\F$-statistic becomes more permissive to parameter mismatches (i.e. the peak 
becomes ``broader'' in the parameter space) at the price of shifting the background
upward~\cite{PhysRevD.93.044058,PhysRevD.97.103020, Dergachev:2019wqa,
Tenorio:2021wmz,Chua:2022ssg}. This same mechanism explains why increasing the coherence
time is a sensible method to increase the significance of a candidate.

\subsection{Single MCMC follow-ups using \pyfstat{}\label{subsec:single-stage}}

The basic idea in~\cite{PhysRevD.97.103020} is that, for a given dataset $x$,
the \emph{posterior probability} on the phase-evolution parameters $\lambda$ can 
be expressed as
\begin{equation}
    p(\lambda | x, \HSignal)  \propto B_{\mathrm{SG}}(x; \lambda) \, p(\lambda) \,,
    \label{eq:posterior}
\end{equation}
where $p(\lambda)$ corresponds to the prior distribution, which in this
context corresponds to the uncertainty on the candidate's parameters produced by a search.
It follows then that an adaptive template bank around the true signal parameters can
be efficiently constructed by sampling the posterior distribution using the semicoherent
$\F$-statistic. In this work, we consider $p(\lambda)$ to be a narrow prior with a width
comparable to the typical parameter-space resolution of a search~\cite{Ashton:2017wui,PhysRevD.104.084012};
this choice has proven successful in the follow-up of candidates from searches using relatively
short coherence times~\cite{PhysRevD.103.064017,KAGRA:2021una,KAGRA:2022osp, LIGOScientific:2021ozr,
LIGOScientific:2022enz,Whelan:2023bha,Tripathee:2023muh}. We refer the reader to~\cite{Covas:2024pam} for
the practicalities of following-up broader parameter-space regions.

As implemented in \pyfstat{}~\cite{PhysRevD.97.103020,Keitel:2021xeq}, the sampling of
Eq.~\eqref{eq:posterior} is performed using \texttt{ptemcee}, a parallel-tempering ensemble 
MCMC sampler~\cite{2013PASP..125..306F,2016MNRAS.455.1919V}, and the \mbox{$\F$-statistic} 
uses the tools provided by LALSuite~\cite{lalsuite,swiglal}. \texttt{ptemcee} 
takes in three configuration parameters: the total number of steps $\Ntot$,
the number of walkers in the ensemble $\Nwalk$, and the number of parallel temperatures 
$\Ntemp$. The total computing cost of a configuration $c = (\Ntot, \Nwalk, \Ntemp)$,
in units of $\F$-statistic evaluations, can be computed as~\cite{PhysRevD.97.103020}
\begin{equation}
    \cost(c) = \Ntot \Ntemp \Nwalk \,.
    \label{eq:cost}
\end{equation}
We postpone to Sec.~\ref{sect:evaluating_effectiveness} the selection of an appropriate
configuration by considering  the false-alarm and false-dismissal probabilities of a 
follow-up stage, which we derive next in Sec.~\ref{subsec:stopping}.

\subsection{Expected sensitivity of a follow-up stage\label{subsec:stopping}}

The sensitivity of a CW all-sky search is typically reported as an estimate of the
required amplitude for a population of isotropically oriented all-sky distributed NSs so
that a significant fraction of them (usually 90\% or 95\%) is deemed
as ``detected''~\cite{PhysRevD.85.042003}.
These populations can be labeled using the sensitivity
depth~\cite{PhysRevD.91.064007,PhysRevD.98.084058}
\begin{equation}
    \depth = \frac{\sqrt{\Sn}}{h_0} \,,
\end{equation}
where $\Sn$ represents the single-sided power spectral density of the noise (PSD).

A single-stage follow-up evaluates the semicoherent
$\F$-statistic on a finite number of templates randomly sampled by an MCMC. 
Upon completion, the $\F$-statistic of the loudest template is compared against
a threshold $\thresh$ to decide whether to discard the candidate or carry it on to the next stage.
If the follow-up is part of a broader set of stages, this veto may depend on results obtained by
previous stages, e.g.~\cite{PhysRevD.104.084012,PhysRevD.106.102008,Ming:2024dug,Steltner:2023cfk,Whelan:2023bha}.
To set up an effective follow-up, it is critical to understand the required detectability thresholds.

The sensitivity of an $\F$-statistic search was thoroughly discussed
in~\cite{PhysRevD.85.042003,PhysRevD.98.084058}. Here we reintroduce
the basic principles and release a new Python code, \texttt{cows3}~\cite{cows3}, 
which implements a version of their semianalytical method. We also extend the
discussion to consider the false-alarm probability associated to an MCMC follow-up 
in Sec.~\ref{subsubsect:pfa}.

\subsubsection{Sensitivity depth at fixed false-dismissal probability\label{subsub:pfd}}

The standard use-case of~\pyfstat{}-like follow-ups~\cite{PhysRevLett.124.191102,
PhysRevD.103.064017,PhysRevD.106.102008,Covas:2022rfg,KAGRA:2021una,KAGRA:2022osp,LIGOScientific:2021ozr,
LIGOScientific:2022enz,Whelan:2023bha,Tripathee:2023muh} calibrates a threshold such that only 
a ``small'' fraction of the signal population is dismissed. This fraction is quantified by means of
the false-dismissal probability $\pfd$, and is a function of both the number of segments 
$\Nseg$ (or coherence time $\Tcoh$) and sensitivity depth of the population. 
Understanding the relation between $\pfd$, $\mathcal{D}$, and $\thresh$ is thus required to
set up an effective follow-up. Throughout this derivation, we assume $\Sn$ has been properly 
estimated~\cite{Krishnan:2004sv,Astone:2005fj,PrixBias} and keep it implicit in our notation. 

The false-dismissal probability associated to a threshold $\thresh$ under $\HSignal$ is given by
\begin{equation}
    \pfd(\thresh, \depth) =
    p(2 \hat{\F} < \thresh | \mathcal{D}) = 
    \int_{-\infty}^{\thresh} \mathrm{d}2\hat{\F}\, 
    p(2 \hat{\F} | \mathcal{D})\,,
    \label{eq:pfd}
\end{equation}
where, to ease the notation we have omitted the number of coherent segments $\Nseg$ and
$\HSignal$ will be replaced by $\depth$ or $\A$ whenever convenient.
Here, $p(2\hat{\F}|\depth)$ represents the $\F$-statistic distribution associated 
to a matched (cf. mismatch discussion in Sec.~\ref{subsubsect:mismatch_vs_pfa})
CW signal drawn from a population at depth $\mathcal{D}$, which is inconvenient 
to evaluate and depends on  the source distribution of amplitude parameters and
sky positions through a marginalization integral
\begin{equation}
    p(2 \hat{\F} | \depth) =
    \int \mathrm{d}\A  \,\mathrm{d}\hat{n} \, 
        p(2 \hat{\F} | \A \, \hat{n}) \, p(\A \, \hat{n} | \depth)\,.
    \label{eq:F_of_D}
\end{equation}
where \mbox{$p(2 \hat{\F} | \A \, \hat{n} \, \depth) =  p(2 \hat{\F} | \A \, \hat{n})$}, 
as $\A$ and $\Sn$ are enough to compute $\depth$.

For a given observing run, the $\F$-statistic depends on the signal's sky position 
$\hat{n}$ and amplitude parameters $\A$ through a deterministic quantity
$\rho^{2}(\A, \hat{n})$, which is usually referred to as the
(squared) signal-to-noise-ratio (SNR, see \cite{PhysRevD.98.084058} for an explicit expression
for $\rho^2$). This allows for Eq.~\eqref{eq:F_of_D} to be expressed as a
one-dimensional integral
\begin{equation}
    p(2 \hat{\F} | \mathcal{D}) = 
    \int_{0}^{\infty} \mathrm{d} \rho^2 p(2 \hat{\F} | \rho^2) p(\rho^2 | \depth) \,,
    \label{eq:F_of_D_SNR}
\end{equation}
where
\begin{equation}
    p(\rho^2|\depth) = \int \mathrm{d} \A  \mathrm{d}\hat{n} \,
    p(\rho^2| \A \, \hat{n}) p(\A \, \hat{n} | \depth)\,
    \label{eq:rho2_of_things}
\end{equation}
and $p(2\hat{\F}| \rho^2) = \chi^{2}_{4\Nseg}(2\hat{\F};\rho^2)$ is a
noncentral chi-squared distribution with $4\Nseg$ degrees of freedom and 
noncentrality parameter
$\rho^2$~\cite{PhysRevD.58.063001,PhysRevD.72.063006}.
Moreover, the sensitivity depth can be factored out of the SNR as
\begin{equation}
    \rho^2(\A, \hat{n}) = \frac{1}{\depth^2} \rho^2_{0}(\A_0, \hat{n}) \,,
\end{equation}
where $\A_0 = \left\{ \cos\iota, \psi \right\}$\footnote{
    $\rho_0^2(\A_0, \hat{n})$, which can be interpreted as a
    ``unit-depth SNR'', will play a similar role to the ``response function''
    $R^2(\theta)$ in~\cite{PhysRevD.85.042003,PhysRevD.98.084058}.
}. 
Equation~\eqref{eq:pfd} is thus an expectation over the distribution of $\rho^2_0$
\begin{equation}
   \pfd(\thresh, \mathcal{D}) = 
   \int_{0}^{\infty} \mathrm{d}\rho^2_{0} \, p(\rho^2_{0})
   p(2 \hat{\F} <  \thresh | \rho^2 = \rho^2_{0} / \depth^2)\,,
   \label{eq:pfd_ez}
\end{equation}
where
\begin{equation}
    p(\rho^2_0) = \int \mathrm{d} \A_{0}  \mathrm{d}\hat{n} \,
    p(\rho^2_0 | \A_{0} \, \hat{n}) \, p(\A_0 \, \hat{n})
    \label{eq:rho2_0}
\end{equation}
and $p(\rho^2_0| \A_{0}\, \hat{n}) = \delta(\rho^2_0 - \rho^2_0(\A_{0}, \hat{n}))$.

The key insight in~\cite{PhysRevD.98.084058,PhysRevD.85.042003} is that,
once the signal population has been specified through the prior  $p(\A_0, \hat{n})$,
Eq.~\eqref{eq:rho2_0} is \emph{independent of the signal's amplitude} and only needs
to be generated \emph{once} per observing run and signal population. As a result, $p(\rho^2_0)$
can be numerically generated and stored as a one-dimensional histogram with which
Eq.~\eqref{eq:pfd_ez} can be easily computed for a given $\thresh$ and
$\depth$~\footnote{
    Incidentally, given a desired false-dismissal probability $\pfd^{*}$,
    we can seamlessly estimate $\depth(\thresh, \pfd^{*})$ by using a root-finding algorithm 
    on Eq.~\eqref{eq:pfd_ez}, as implemented in~\cite{octapps}.
    We provide the first open-source Python implementation of this
    algorithm in~\cite{cows3}.
}.

\begin{figure}
    \centering
    \includegraphics[width=\columnwidth]{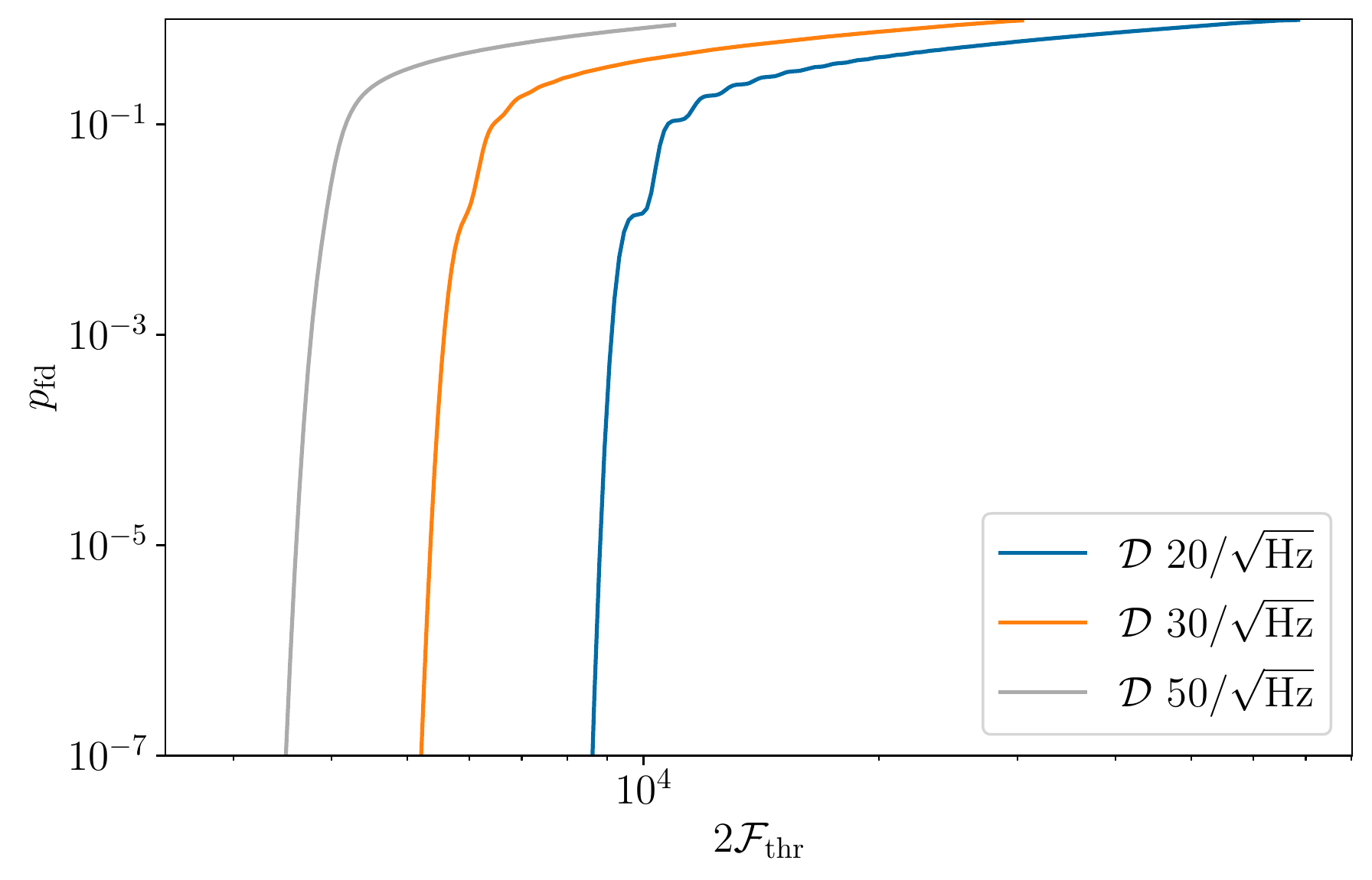}
    \caption{
        False-dismissal probabilities [Eq.~\eqref{eq:pfd_ez}]
        for a simulated 1-year observing run with two detectors compatible
        with H1 and L1 using $\Nseg = 730$ ($\Tcoh=\SI{0.5}{\day}$) for 
        three representative sensitivity depths.
    }
    \label{fig:Pfd_vs_2F}
\end{figure}

In Fig.~\ref{fig:Pfd_vs_2F} we compute the false-dismissal probability
using Eq.~\eqref{eq:pfd_ez} for three representative depth values,
namely $\depth = \{20, 30, 50\} / \sqrt{\unit{\hertz}}$.
These cover the typical range of sensitivities achieved by all-sky
searches with $\Tcoh \lesssim 0.5\,\mathrm{days}$~\cite{PhysRevD.103.064017,
PhysRevD.106.102008, PhysRevD.100.024004,PhysRevD.96.062002,LIGOScientific:2018gpj,
Covas:2022rfg,Covas:2024nzs}. 
We simulated a 1-year observing run with two detectors consistent with the
Advanced LIGO  detectors~\cite{LIGOScientific:2014pky}, LIGO Hanford (H1)
and LIGO Livingston (L1), starting at the beginning O3. 
Threshold values are computed using $\Nseg=730$
(i.e.~$\Tcoh=\SI{0.5}{\day}$), which is a standard choice for a first-stage
follow-up~\cite{KAGRA:2021una,PhysRevD.103.064017,PhysRevD.105.022002,
KAGRA:2022osp, Whelan:2023,LIGOScientific:2022enz,PhysRevD.106.102008}.

Generally, single-stage follow-ups estimate $\pfd$ as the fraction of dismissed signals
in an simulated injection campaign; as a result, for a given search,
$\pfd$ is lower-bounded by the number of simulated signals. 
By means of Eq.~\eqref{eq:pfd_ez}, on the other hand, we can compute
arbitrarily low false-dismissal probabilities (insofar the signal model in use matches
the expected signal). Typical follow-ups operate 
at~\mbox{$\pfd \in [10^{-5}, 10^{-3}]$}~\cite{PhysRevLett.124.191102,PhysRevD.103.064017,
PhysRevD.106.102008,Covas:2022rfg}. In this work, we chose a conservative value of
\mbox{$\pfd = 10^{-5}$} to present our results.

\subsubsection{A note on $\pfa$\label{subsubsect:pfa}}

Given a threshold $\thresh$, the corresponding false-alarm probability $\pfa$ can be evaluated as 
the survival function of the detection statistic at hand under the noise hypothesis
\begin{equation}
    \pfa(\thresh) = \int_{\thresh}^{\infty} \mathrm{d} 2\hat{\F} \, p(2 \hat{\F}|\HGauss)\,.
\end{equation}
This quantity could be interpreted as the fraction of nonastrophysical CW candidates
that would not be discarded by a follow-up stage assuming Gaussian noise. Note, however,
that real searches are affected by non-Gaussian noise and disturbances implying that a larger fraction of
non-CW-related outliers might pass this first selection. Moreover, we here neglect the fact that
selected candidates in a follow-up correspond to the \emph{loudest} candidates of a previous search step.

The distribution of the $\F$-statistic under the Gaussian noise hypothesis is well-known and corresponds
to a central chi-squared distribution with $4 \Nseg$ degrees of
freedom~\cite{PhysRevD.58.063001,PhysRevD.72.063006}.  In a follow-up, 
similarly to some searches (e.g.~\cite{Keitel:2019zhb,Wette:2021tbv,distromax,LIGOScientific:2021quq}), 
the statistic upon which a threshold is imposed is actually the \mbox{$\F$-statistic} of the
\emph{loudest} template, $2\hat{\F}_{\mathrm{max}}$. 
The distribution of such statistic was discussed in~\cite{PhysRevD.104.084012, distromax}, 
and corresponds to a Gumbel distribution
\begin{equation}
    p(2\hat{\F}_\mathrm{max} |\HGauss) = 
    \frac{1}{\sigma_{\mathrm{G}}}
    e^{
        -\left(\frac{2\hat{\F}_{\mathrm{max}}- \mu_{\mathrm{G}}}{\sigma_{\mathrm{G}}}\right) 
        - e^{-\left(\frac{2\hat{\F}_{\mathrm{max}} -\mu_{\mathrm{G}}}{\sigma_{\mathrm{G}}}\right)}
    }\,,
    \label{eq:Gumbel}
\end{equation}
where the location $\mu_{\mathrm{G}}$ and scale $\sigma_{\mathrm{G}}$ parameters can be numerically
characterized. As discussed in~\cite{distromax}, Eq.~\eqref{eq:Gumbel} is generally applicable in the 
presence of non-Gaussianities, such as spectral lines~\cite{Covas:2024pam,Jaume:2024eob}.

It was first proposed in~\cite{PhysRevD.104.084012} to use off-sourcing~\cite{Isi:2020uxj} to re-evaluate
the template bank produced by an MCMC follow-up at different sky points and get the loudest point
to numerically generate $p(2\hat{\F}_{\mathrm{max}}|\HGauss)$.
During the development of this work,
we noted that such an approach \emph{underestimates} the expected background of an MCMC.
This is related to the fact that MCMC sampling tends to favor high-likelihood regions,
as opposed to a random draw in which different templates are drawn independently.
The loudest value of an MCMC, thus, \emph{tends to be higher} than the loudest of a random template bank\footnote{
    This has no effect on past results, as every application calibrated its own follow-up.
    On the other hand, this result explains the unusually high Bayes factors quoted
    in~\cite{PhysRevD.104.084012}, where the noise distribution was severely underestimated.
}.

\begin{figure}
    \includegraphics[width=\columnwidth]{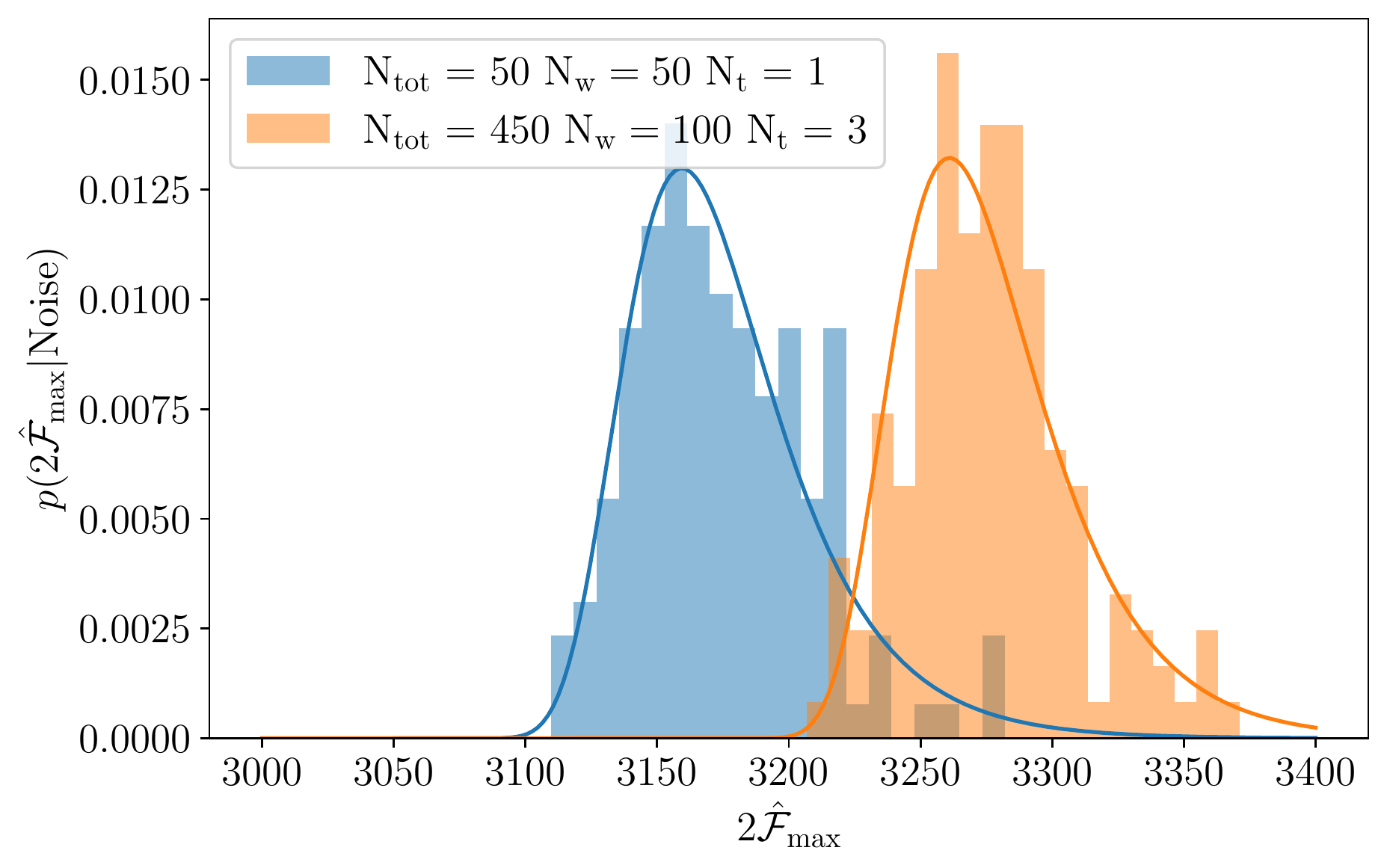}
    \caption{
        Noise distributions for two particular follow-up configurations, as specified in the legend.
        Each histogram is generated using 150 Gaussian noise realizations of 1-year observing runs
        with 2 detectors using $\Nseg = 730$. The solid lines represent the best fit for a Gumbel
        distribution using the method of \texttt{scipy.stats.gumbel\_r}~\cite{scipy}.
        Results are obtained using the priors for a search of isolated emitters in Table~\ref{tab:inj_pars_and_prior} with width calculated for $\Tcoh=\Tsft$.
    }
    \label{fig:numerical_pfa}
\end{figure}

In Fig.~\ref{fig:numerical_pfa}, we numerically generate the noise distribution of two particular
MCMC configurations with the priors listed in Table~\ref{tab:inj_pars_and_prior}
for a search of isolated emitters.  We generated 150 Gaussian noise realizations with
the same data setup as in Fig.~\ref{fig:Pfd_vs_2F}.
This number of realizations allows us to estimate false-alarm probabilities down to
$\pfa \gtrsim 10^{-2}$,  which is enough for a first-stage follow-up
(see e.g.~\cite{PhysRevD.106.102008}). We observe that the general behavior of 
the distribution is well captured by a Gumbel distribution, as discussed in~\cite{distromax}.
We also note that the specific parameters  of the Gumbel distribution are a function of the 
chosen configuration. To draw fair conclusions, we estimate the Gumbel parameters for a 
relevant subset of the MCMC considered in this work. The results are collected in 
Appendix~\ref{app:noise_distribution}.

\subsubsection{Mismatch corresponds to false-alarm probability\label{subsubsect:mismatch_vs_pfa}}

Throughout Sec.~\ref{subsub:pfd} we implicitly assumed the phase-evolution
parameters of a candidate were perfectly matched by the follow-up's template bank,
so that the recovered SNR corresponded to the actual SNR of the signal. In practice,
however, the signal's parameters $\lambda_{\mathrm{s}}$ will be off by
$\Delta \lambda$ from the closest template in the bank. This deviation is quantified by
\emph{mismatch}, which is defined as the fractional SNR loss due to
$\Delta \lambda$~\cite{PhysRevD.94.122002}
\begin{equation}
    \mu(\Delta \lambda; \lambda_{\mathrm{s}}) =
    \frac{
        2\hat{\F}(\lambda_{\mathrm{s}}) - 2\hat{\F}(\lambda_{\mathrm{s}}+\Delta\lambda)
    }{
        2\hat{\F}(\lambda_{\mathrm{s}}) - 4 \Nseg
    }\,,
    \label{eq:mismatch}
\end{equation}
where $4\Nseg$ corresponds to the expected $2\hat{\F}$ value in noise.

\begin{figure}
    \includegraphics[width=\columnwidth]{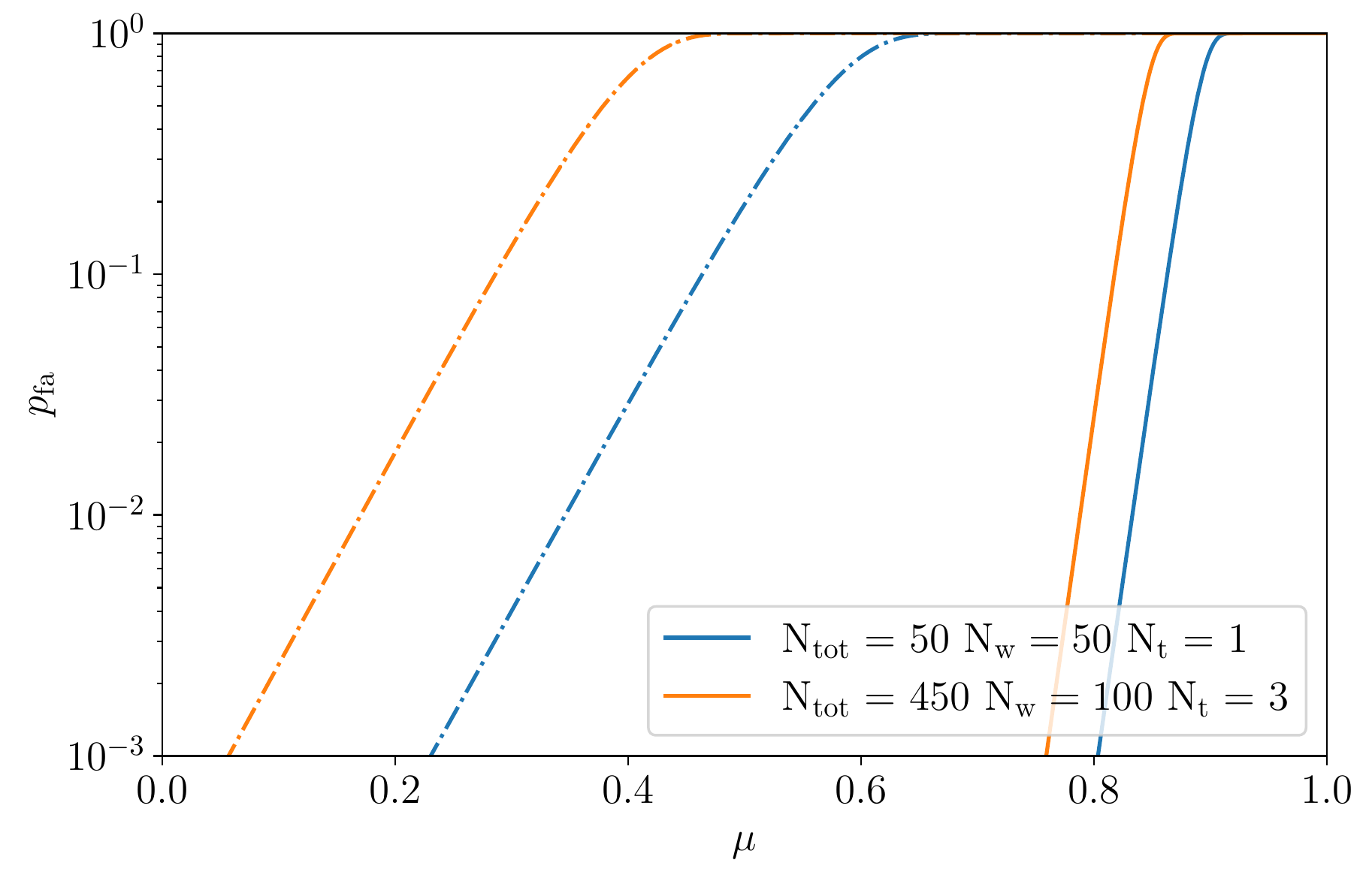}
    \caption{
        False-alarm probability as a function of mismatch for the two MCMC configurations
        shown in Fig.~\ref{fig:numerical_pfa}. Signal populations are compatible with a population
        of isolated neutron stars at~\mbox{$\depth = 30 /\sqrt{\mathrm{Hz}}$} (solid lines)
        and~\mbox{$\depth = 50 /\sqrt{\mathrm{Hz}}$} (dash-dotted lines).
    }
    \label{fig:Pfa_vs_mu}
\end{figure}

To account for a mismatch $\mu$, the threshold $\thresh$  at a fixed $\pfd$
must be \emph{lowered} to
\begin{equation}
    \thresh(\mu) = \thresh \cdot (1 - \mu) + 4 \Nseg \mu \,,
    \label{eq:thres_mu}
\end{equation}
which corresponds to a reduction from the optimal SNR $\rho^2$ to
$\rho^2 (1 - \mu)$.  This implies that, at a given $\pfd$, 
the false-alarm probability of the follow-up is now a function of mismatch
\begin{equation}
    \pfa (\mu) = \int_{\thresh(\mu)}^{+\infty} \mathrm{d} 2\hat{\F}  \,
    p(2\hat{\F}_{\mathrm{max}} |\HGauss) \,.
    \label{eq:pfa}
\end{equation}
We show the resulting false alarm probability using the distributions discussed
in Sec.~\ref{subsubsect:pfa} and Appendix~\ref{app:noise_distribution} in
Fig.~\ref{fig:Pfa_vs_mu}. Depending on sensitivity depth, acceptable 
mismatches (i.e. $\pfa(\mu) \approx 10^{-2}$) are as high as
$\mu \approx 0.8$.

The result in Eq.~\eqref{eq:pfa} culminates in the proposal of mismatch
$\mu$ as a metric for the effectiveness of a follow-up: 
\emph{any} signal can be ``detected'' by a follow-up
if enough false-alarm probability is allowed (i.e. a low-enough threshold is used).
The goal of an effective follow-up is to claim said ``detection''
in such a way that \emph{a significant fraction of the noise candidates} is discarded
in a single stage. Failing to operate at a low false-alarm probability defeats the purpose
of a fast follow-up, as computing cost is then simply pushed forward to further stages.

Note that the $\pfa$ here computed is a simple construct to compare the effectiveness of different
configurations on equal grounds. This is particularly important in the case of a follow-up,
as the selected candidates are usually not a random selection but rather the \emph{top-scoring} templates 
from a previous search stage, thus implying our $\pfa$ could fall short on a real search case. However, the typical difference in segment lengths between a search such as the ones
here considered ($N_{\mathrm{seg}}\sim 10^{4}$) and a follow-up ($N_{\mathrm{seg}} \sim 10^3$) is
likely to diminish the effects of such choice. Since follow-ups tend to operate by selecting candidates based
on computing cost and not false-alarm probability, we leave such a discussion for future work and limit the usage
of $\pfa$ to the comparison of different MCMC configurations.

\subsection{A new coherence-time ladder\label{subsect:ladder}}

As a final step, we address the setup of a coherence-time ladder. As described
in~\cite{Ashton:2017wui,PhysRevD.104.084012}, this is a multistage follow-up setup 
in which coherence time (number of segments) is gradually increased (decreased) in 
order to impose stronger constraints and zoom in on CW candidates.
This can be seen as a rudimentary form of sequential Monte 
Carlo~\cite{smc}.

The standard construction of the ladder~\cite{Ashton:2017wui} uses the parameter-space metric, which is only
known analytically for a limited number of cases~\cite{PhysRevD.57.2101,Brady:1998nj,Jaranowski:1998ge,
PhysRevD.75.023004,Pletsch:2008gc,Pletsch:2010xb,PhysRevD.88.123005,PhysRevD.90.122010,
PhysRevD.91.102003,Wette:2015lfa}; for more general cases, a numerical approach is required,
as discussed in~\cite{Covas:2024pam}.  Here, we propose a simpler approach using the scaling of
individual parameter-space dimensions with $\Tcoh$.

The ladder construction goes as follows: suppose we conduct a follow-up on a parameter space region 
$\Delta \lambda^{(j)}$ using a coherence time $\Tcoh^{(j)}$. The number of ``independent templates'' in said
region can be estimated as
\begin{equation}
    \N(\Delta \lambda^{(j)}, \Tcoh^{(j)}) = 
    \displaystyle\int_{\Delta \lambda^{(j)}} \mathrm{d}\lambda \, \sqrt{g\left(\Tcoh^{(j)}\right)} \,,
    \label{eq:Ntemplate}
\end{equation}
where $g$ is the determinant of the parameter-space metric.
It was argued in~\cite{Ashton:2017wui} that \pyfstat{} is effective for parameter-space
regions  containing $\N_{*}$ templates, typically around $10^{3}$ to $10^{4}$.
This was reinterpreted in~\cite{PhysRevD.104.084012} as the fraction of volume down to which 
a follow-up can successfully compress a posterior distribution. Broader regions were later
explored in~\cite{Covas:2024pam}.

After a successful follow-up stage, the resulting region 
$\Delta \lambda^{(j+1)}$ will contain a significantly smaller number of templates than the original one,
\mbox{$\mathcal{N}(\Delta \lambda^{(j+1)}, \Tcoh^{(j)}) \sim 1$}.  Also, in this work we consider follow-ups as 
local analyses, which means the number of templates can be approximated as
\begin{equation}
    \N(\Delta \lambda^{(j)}, \Tcoh^{(j)}) \approx \mathrm{Vol}(\Delta \lambda^{(j)}) \sqrt{g\left(\Tcoh^{(j)}\right)} \,.
\end{equation}
The new coherence time $\Tcoh^{(j+1)}$, thus, should be chosen such that
\begin{equation}
    \begin{split}
    \N_{*} 
    &= \N(\Delta \lambda^{(j+1)}, \Tcoh^{(j+1)}) \\
    &\approx 
        \frac{\mathcal{N}\left(\Delta \lambda^{(j+1)}, \Tcoh^{(j+1)}\right)}{\mathcal{N}
        \left( \Delta \lambda^{(j+1)}, \Tcoh^{(j)}\right)} 
    \approx \frac{\sqrt{g\left(\Tcoh^{(j+1)}\right)}}{\sqrt{g\left(\Tcoh^{(j)}\right)}}\,.
    \end{split}
    \label{eq:ladder}
\end{equation}
where to go to the second line we used $\mathcal{N}(\Delta \lambda^{(j+1)}, \Tcoh^{(j)}) \sim 1$
to introduce a term containing $\mathrm{Vol}(\Delta \lambda^{(j)})$ to cancel it from the numerator,
as described in~\cite{Ashton:2017wui}. An alternative interpretation is given in~\cite{PhysRevD.104.084012}.

Equation~\eqref{eq:ladder} requires to evaluate the determinant of the parameter-space metric.
Since the follow-up is a local analysis, any dependency on the parameter-space position is likely
to be negligible and coherence time $\Tcoh$ will be the only quantity that changes across stages.
We also note that, for semicoherent searches, the resolution of relevant phase-evolution
parameters tends to scale with
$\Tcoh^{-1}$~\cite{Krishnan:2004sv,PhysRevD.75.023004,PhysRevD.91.102003}.
These two facts motivate the following ansatz to estimate the number of templates in
a follow-up stage
\begin{equation}
    \N_{*} = \left(\frac{\Tcoh^{(j+1)}}{\Tcoh^{(j)}}\right)^{D} \,.
\end{equation}
Here, $D$ is the number of resolved parameters considered by the follow-up.
Therefore, the coherence-time ladder can be constructed as easily as
\begin{equation}
    \Tcoh^{(j)} = \N_{*}^{j/D} \, \Tcoh^{(0)}\,.
    \label{eq:ladder_approx}
\end{equation}
Choosing an appropriate value of $D$ is not trivial in general, as different parameter-space dimensions
will have different dependencies on $\Tcoh$ depending on typical variation scale of the signal's phase
(see e.g.~the short-segment and long-segment metrics in~\cite{PhysRevD.91.102003}).
In this work, we choose
$D$ assuming that all the resolvable parameters scale with $\Tcoh^{-1}$. 
This returns a higher value than required, and thus allows us to operate in the safe side.
Since we are dealing with isolated signals with one spindown term and circular binaries,
we choose $D=4$ for the isolated case and $D=6$ for the binary case.
Other signal models, such as higher number of spindowns or eccentricity in the binary orbit,
may require extra stages, depending on the chosen configuration, to ensure proper convergence.

\section{How to evaluate the effectiveness of CW follow-ups\label{sect:evaluating_effectiveness}}

We are interested in follow-up configurations with 1) a low computing cost and 2) a low false-alarm probability
at a fixed false-dismissal probability and sensitivity depth. This is equivalent to a multiobjective optimization
problem, which is in general solved by a \emph{family of equivalent solutions}, each of them representing a 
different trade-off between 1) and 2)~\cite{Emmerich2018}. As per Sec.~\ref{sect:follow_up_introduction}, 
rather than minimizing $\pfa$ an effective follow-up can be defined so that $\pfa$ is below a \emph{maximum}
allowed value; using Eq.~\eqref{eq:pfa}, this corresponds to a maximum allowed mismatch $\mu_{\mathrm{M}}$.
As a result, we define the optimal follow-up configuration $c_*$ so that
\begin{equation}
    c_*  = \argmin_{c}  \cost(c) \,, \mu(c) < \mu_{\mathrm{M}} \,,
\end{equation}
where $\cost(c)$ is computed using Eq.~\eqref{eq:cost} and $\mu(c)$ is the mismatch
associated to a follow-up configuration $c$.  

The considered list of follow-up configurations is shown in Table~\ref{tab:MCMC_setups}.
We detail the computation of $\mu_{\mathrm{M}}$ in Sec.~\ref{sec:max_mismatch} and
$\mu(c)$ in Sec.~\ref{sec:configuration_mismatch}. This process makes use of a campaign of injected 
software-simulated signals described in Sec.~\ref{subsec:campaign}. Results are presented in 
Sec.~\ref{sect:optimizing_setup}.

\subsection{Maximum allowed mismatch\label{sec:max_mismatch}}

As discussed in Sec.~\ref{subsec:stopping}, for a given population depth $\depth$ and false-dismissal
probability  $\pfd$, and for the case \mbox{$\mu = 0$}, we can construct a detectability threshold
$\thresh$ by numerically inverting Eq.~\eqref{eq:pfd}.
To account for a mismatch $\mu$, this threshold should be modified using Eq.~\eqref{eq:thres_mu}
into $\thresh(\mu)$. Finally, Eq.~\eqref{eq:pfa} links the detectability threshold $\thresh(\mu)$
to the resulting false-alarm probability $\pfa(\mu, c)$.\footnote{Note that this last step depends on $c$ through
$p(2\hat{\F}_{\mathrm{max}}|\mathcal{H}_{\mathrm{G}})$.} This establishes a monotonic 
relation between $\pfa$ and $\mu$ (for a given $\pfd$, $\depth$, and $c$) which can be numerically 
inverted to compute \mbox{$\mu_{\mathrm{M}} = \mu(\pfa, c)$}.

We consider the case $\pfd = 10^{-5}$ and a range of false-alarm probabilities $\pfa \in \left[10^{-2}, 10^{-1}\right]$.
This produces a range of $\mu_{\mathrm{M}}$ values for each configuration $c$. To evaluate our results, we consider
the broadest range of $\mu_{\mathrm{M}}$ across all configurations at a fixed cost, which we show 
as shaded areas in Figs.~\ref{fig:mismatch_vs_cost} and~\ref{fig:mismatch_vs_cost_real_data}.
Acceptable follow-up configurations are lower than the shaded area.

\subsection{Mismatch of a configuration\label{sec:configuration_mismatch}}

To associate a mismatch to a configuration $c$, we use a software-simulated injection campaign,
described in Sec.~\ref{subsec:campaign}. For each injection, we run an MCMC follow-up
using the configuration $c$ and retrieve the loudest \mbox{$\F$-statistic}.
This value is compared to the \mbox{$\F$-statistic} at the injection point to compute 
the resulting mismatch [Eq.~\eqref{eq:mismatch}]. The result is a sample of mismatches,
one for each injection,  which together describe the mismatch distribution associated
to the chosen MCMC configuration $c$.  We then take the $90^{\mathrm{th}}$ quantile of this sample as
a proxy for the mismatch of the configuration \mbox{$\mu(c) = \muq(c)$}.
This last choice is motivated by the available computing resources; a subset of the configurations was re-evaluated 
on a larger injection campaign using the $95$th and $99$th percentile, with consistent results. 
All the injections used in this procedure return an $\F$-statistic above the detectability thresholds associated
to the probed values of $\pfd$ and $\pfa$ in the first follow-up stage.

We also compute the track distance~\cite{PhysRevD.103.064053}
\begin{equation}
        d(\lambda, \lambda_{\rm s})= \Tsft \langle \left| f(t;\lambda) - f(t;\lambda_{\rm s})\right| \rangle \,,
        \label{eq:distance}
\end{equation}
which evaluates the distance between two templates as the average area enclosed between
their corresponding frequency tracks on the spectrogram. The notation $\langle \cdot \rangle$
denotes averaging along a subset of the timestamps of the observing run,
as discussed in~\cite{PhysRevD.103.064053}. This will allow us to assess whether the loudest candidate
from an MCMC follow-up corresponds to the injected signal or a noise fluctuation.

\begin{table}
    \centering
\begin{tabular}{lc}
        \toprule
        $\Nwalk$ & 50, 100 \\
        $\Ntemp$ & 1, 2, 3 \\ 
        $\Ntot$ & 50, 100, \dots, 450, 500\\
        \bottomrule
    \end{tabular}
    \caption{
        Parameter space of the MCMC configurations considered in this work.
        Each of the 54 configurations  $c$ is an element of the cartesian product
        of these three parameters. $\Ntot$ is varied in steps of 50.
    }
    \label{tab:MCMC_setups}
\end{table}

\subsection{Injection campaign\label{subsec:campaign}}

We simulate two injection campaigns of isotropically oriented all-sky CW sources
in Gaussian and real noise in a consistent manner with the two all-sky CW searches
produced by the LVK in O3a and O3 data~\cite{PhysRevD.103.064017,PhysRevD.106.102008}.
The resulting MCMC follow-up performances will thus be informative to future all-sky searches. 
In both cases, we consider two detectors in a location consistent with the Advanced LIGO
detectors. 

We assume a PSD of \mbox{$\sqrt{\Sn} = 10^{-23} / \sqrt{\mathrm{Hz}}$} for gapless Gaussian noise.
O3 data is generated using the science segments and gating times available in~\cite{science_segments,gating}. 
The injection parameter spaces are summarized in Table~\ref{tab:inj_pars_and_prior}. Frequency and binary orbital
parameters are drawn uniformly in the specified ranges; sky positions are drawn uniformly across the celestial sphere;
spindown values are drawn from a mixture of two log-uniform distributions with relative weights of 0.9 for the negative
range and 0.1 for the positive range; the cosines of the inclination angles, the polarization angles,
and the initial phase are drawn  from a uniform distribution along 
$[-1, 1]$,  $[-\pi/4, \pi /4]$ and $[0, 2\pi]$, respectively.

For sources in binary systems, we use \mbox{$\Tobs = 0.5~\mathrm{year}$} and generate 
50\%-overlapping Short Fourier Transforms (SFTs)~\cite{sft} with a duration
of \mbox{$\Tsft = \SI{1024}{\second}$} to be consistent with~\cite{PhysRevD.103.064017}.
For isolated sources, we use \mbox{$\Tobs = 1~\mathrm{year}$} and nonoverlapping
\mbox{$\Tsft = \SI{1800}{\second}$} SFTs.
Signals are injected at two representative sensitivity depths, namely
$30$ and $50/\sqrt{\mathrm{Hz}}$ for isolated and $20$ and $25/\sqrt{\mathrm{Hz}}$
for sources in binary systems. This discrepancy between isolated and binary sources arises 
from the different parameter-space dimensionality.

For completion, the cost of a single $\F$-statistic evaluation, which depends on the number
of SFTs~\cite{CFSv2}, is about $\SIrange{0.4}{0.7}{\second}$ in a 13th Gen. Intel(R) Core(TM) i7-1355U. 

\begin{table}[!bh]
    \centering
    \renewcommand*{\arraystretch}{1.3}
    \begin{tabular}{ccc}
        \toprule
         & Interval & Prior width \\
         \midrule
        $f_0$ & [50, 500] Hz & $\frac{1}{2\Tcoh}$\\
        $f_1$ & [$-10^{-8}$, $10^{-9}$] Hz/s & $\frac{1}{2 \Tcoh \Tobs}$\\
        $\alpha$ & [$-\pi$, $\pi$] rad & $\frac{10^4}{2 f \Tcoh}$\\
        $\delta$ & [$-\pi/2$, $\pi/2$] rad & $\frac{10^4}{2 f \Tcoh}$\\
        $P$ & [7, 15] d &$\frac{\pi}{f\, a_p \Tcoh\Tobs \Omega^3 }$\\
        $a_p$ & [5, 15] lt-s & $\frac{1}{2 f \Tcoh \Omega}$\\
        $\tasc$ & [$t_\mathrm{mid}-\frac{P}{2}$, $t_\mathrm{mid}+\frac{P}{2}$] & $\frac{5}{2 f\,a_p\,\Tcoh\,\Omega^2}$\\
        \bottomrule
    \end{tabular}
    \caption{
        Parameter-space covered by the injection campaigns.
        The prior's width corresponds to the standard deviation of a Gaussian prior, 
        except for $\tasc$ which uses a uniform prior. 
        $\tmid$ refers to the timestamp at the middle of the observing run.
        For the first follow-up stage, we calculate the widths using \mbox{$\Tcoh = \Tsft$}.
    }
    \label{tab:inj_pars_and_prior}
\end{table}

\section{Optimizing the choice of follow-up setup\label{sect:optimizing_setup}}

 \begin{figure*}
    \centering
    \includegraphics[width=\columnwidth]{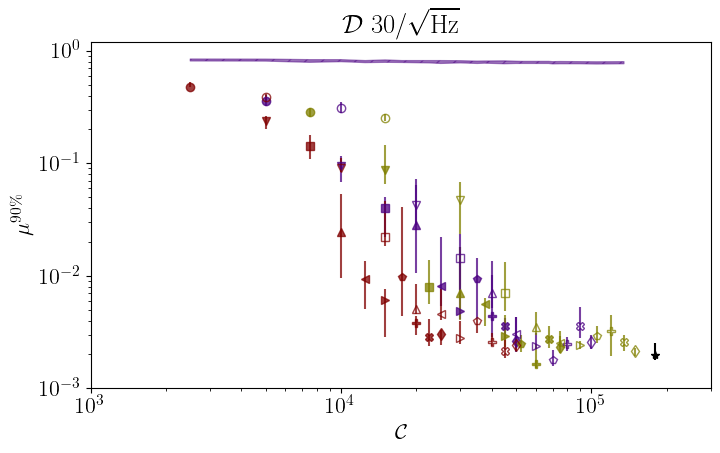}
    \includegraphics[width=\columnwidth]{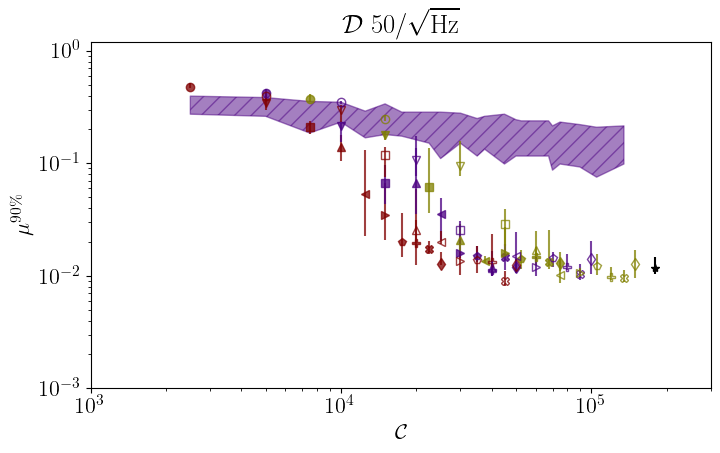}
    \includegraphics[width=\columnwidth]{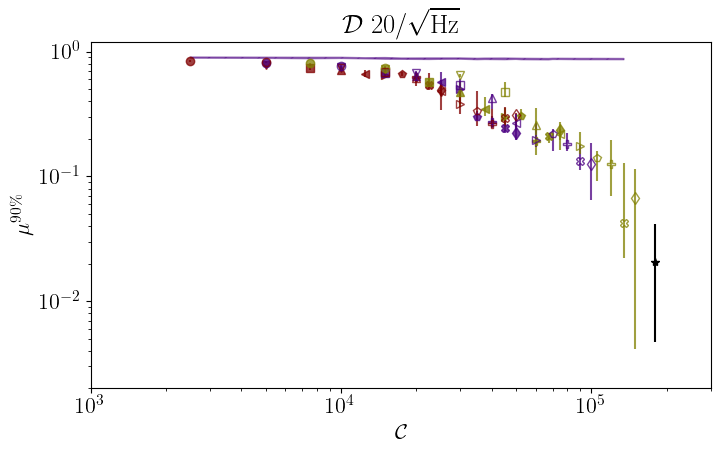}
    \includegraphics[width=\columnwidth]{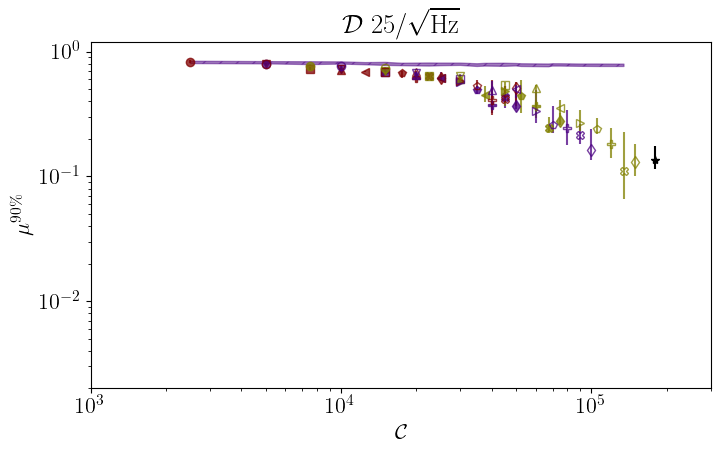}
    \caption{
        $90$th percentile of the $\mu$ distribution as a function of the cost of the MCMC configuration
        using $\Tcoh=\SI{0.5}{\day}$. 
        The upper row corresponds to the isolated injection campaign ($\Tobs=1$~year),
        while lower row corresponds to  the binary injection campaign ($\Tobs=0.5$~year).
        The shaded regions correspond to $\mu$ values such that  $10^{-2}\leq\pfa\leq10^{-1}$
        and $\pfd=10^{-5}$ at the specified sensitivity depth (see Sec.~\ref{subsubsect:pfa} and Appendix~\ref{app:noise_distribution}). 
        Each configuration's performance is identified by a different marker.
        The starred black marker represents the performance of the standard configuration used in previous searches (see text).
        The black triangular marker
        denotes the configuration \mbox{$\Nwalk=50,\Ntemp=1, \Ntot=300$} that we identify as a good general choice.
        Filled (unfilled) markers correspond  to $N_w=50 \, (100)$.
        Red, purple, and green markers correspond to $N_t=1,\,2,\,3$, respectively.  The shape of the marker denotes different
        values of $N_{\mathrm{tot}}$. Uncertainties are reported at 68\% confidence level using bootstrapping.
        Any configuration below the shaded region is appropriate to successfully follow up a signal at the specified
        depth level.
        }
    \label{fig:mismatch_vs_cost}
\end{figure*}

We present the results of evaluating  the 54 different MCMC configurations
listed in Table~\ref{tab:MCMC_setups} on the injection campaigns described
in Sec.~\ref{subsec:campaign}. Additionally, we run an extra MCMC configuration
corresponding to the standard choice in previous MCMC 
follow-ups~\cite{KAGRA:2021una,PhysRevD.103.064017,PhysRevD.105.022002,KAGRA:2022osp,
Whelan:2023,LIGOScientific:2022enz,PhysRevD.106.102008}
given by $\Nwalk = 100, \Ntemp = 3, \Ntot = 600$.

We use uncorrelated Gaussian priors with a width given by the typical parameter
resolution (see Table~\ref{tab:inj_pars_and_prior}). This is consistent with a maximum-entropy
distribution given the average location and uncertainty of a parameter~\cite{Jaynes:2003jaq}.
To mimic the behavior of an actual search, we randomly shift the priors' centre away
from the injection point. To do so, we draw the priors' center from the Gaussian 
(uniform for $t_{\rm asc}$)  prior centered at the injection  parameters with a 
standard deviation as reported in 
Table~\ref{tab:inj_pars_and_prior}~\cite{Krishnan:2004sv, PhysRevD.103.064017, PhysRevD.91.102003}.

We report the results of our optimization in Sec.~\ref{subsect:stage-0} for a single-stage 
follow-up and Sec.~\ref{subsect:multi-stage} for multistage follow-up. General discussion
is presented in Sec.~\ref{subsect:discussion}.

\subsection{Single-stage follow-up\label{subsect:stage-0}}

The first follow-up stage uses a coherence time of $\Tcoh = \SI{0.5}{\day}$,
in a similar manner to the O3 all-sky searches~\cite{PhysRevD.104.084012, PhysRevD.106.102008}.
This is a significant increase with respect to the searches' coherence time ($\Tcoh \sim \Tsft$),
and has been empirically verified to successfully find simulated CW signals given the uncertainties of
a typical all-sky search both in real-data searches~\cite{PhysRevLett.124.191102, 
PhysRevD.103.064017,PhysRevD.106.102008,Covas:2022rfg,KAGRA:2021una,KAGRA:2022osp,LIGOScientific:2021ozr,
LIGOScientific:2022enz,Whelan:2023bha,Tripathee:2023muh} and in broader follow-up studies~\cite{Covas:2024pam}.

Note that this starting value is not given by the coherence-time ladder. This is because previous search stages
using shorter coherence times are likely to use different detection statistics that behave differently to the
\mbox{$\F$-statistics}; as a result, the first stage is empirically accommodated for the typical uncertainties
returned by said stages.

As discussed in Sec.~\ref{subsubsect:pfa}, we are interested in configurations to find signals with a mismatch compatible
with  $\pfa\lesssim 10^{-2}$ and $\pfd \lesssim 10^{-5}$ at the specified sensitivity depths.
To do so, for each follow-up configuration, we calculate $\muq$ and the track distance [Eq.~\eqref{eq:distance}]
using 100 simulated signals. These results in simulated data are shown in Fig.~\ref{fig:mismatch_vs_cost}
for both the isolated and binary cases.

\begin{figure*}
    \centering
    \includegraphics[width=\columnwidth]{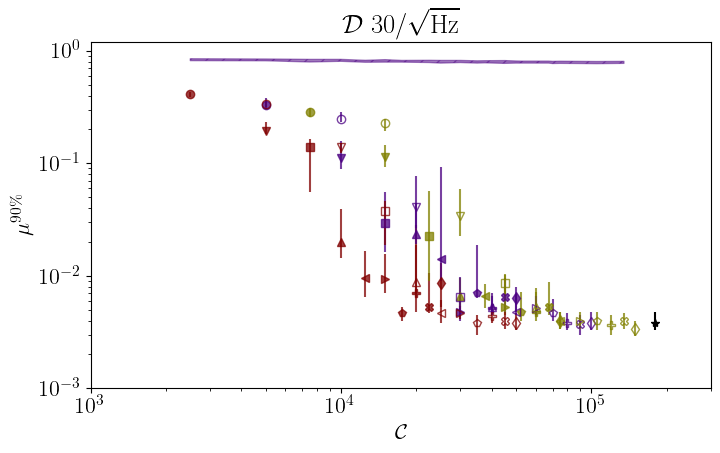}
    \includegraphics[width=\columnwidth]{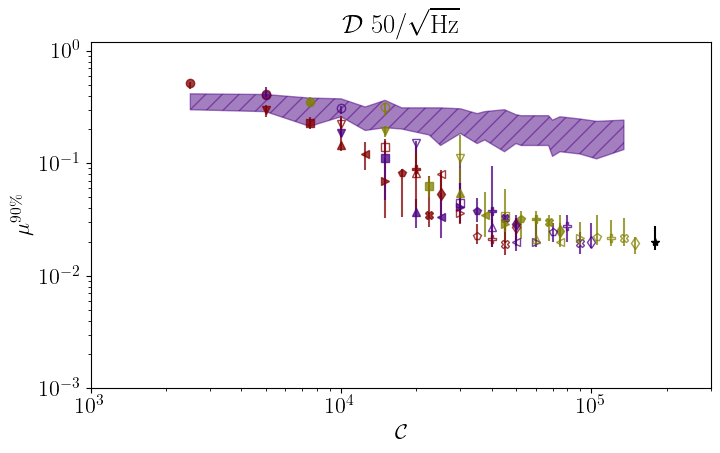}
    \includegraphics[width=\columnwidth]{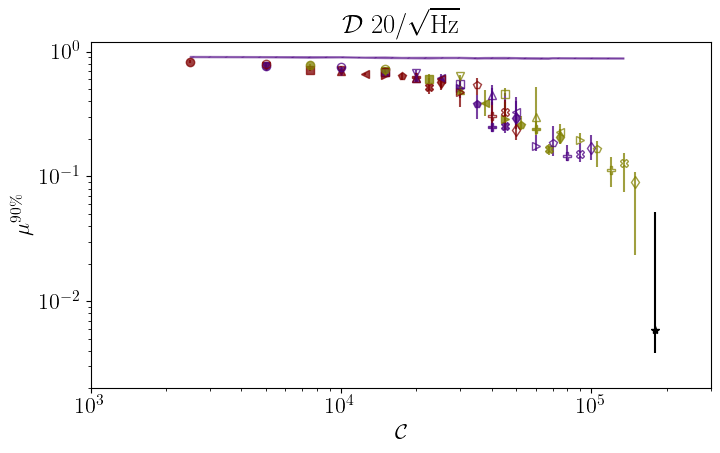}
    \includegraphics[width=\columnwidth]{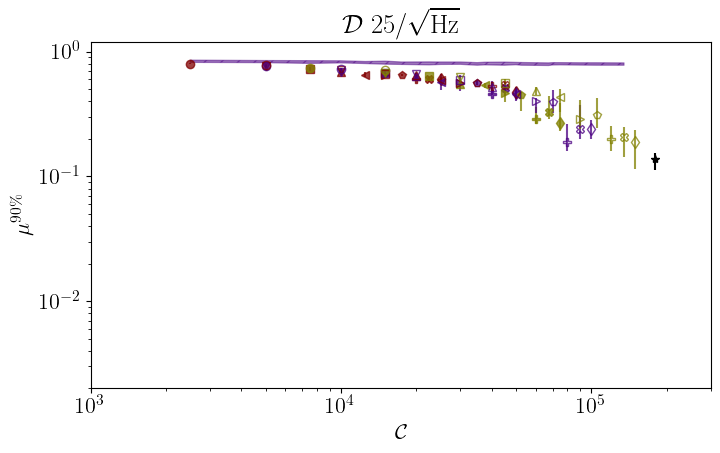}
    \caption{Same results as in Fig.~\ref{fig:mismatch_vs_cost} but using Advanced LIGO O3 and O3a data for isolated
    (upper row) and binary (lower row) signals, respectively.}
    \label{fig:mismatch_vs_cost_real_data}
\end{figure*}

To test the robustness of these configurations to deviations from Gaussianity in real noise,
the same configurations are evaluated by injecting the same simulated signals in real O3 Advanced LIGO data.
The results are shown in Figs.~\ref{fig:mismatch_vs_cost_real_data}.
We observe no significant deviation with respect to the Gaussian-noise injection campaign,
demonstrating the capabilities of MCMC follow-ups to identify signals in a realistic situation.
Note that we still use of the results from Appendix~\ref{app:noise_distribution} 
using Gaussian noise, meaning the false-alarm probability does not correspond to $\pfa(\mu, c)$.
As we will see in Sec.~\ref{sect:impact}, this is not a problem, as real-data searches tend to operate 
by selecting a specific number of candidates regardless of the actual $\pfa$. 

\begin{figure*}
    \centering
    \includegraphics[width=\columnwidth]{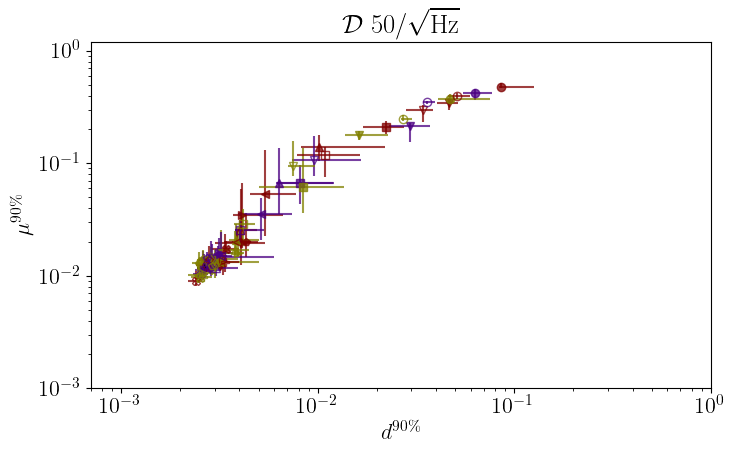}
    \includegraphics[width=\columnwidth]{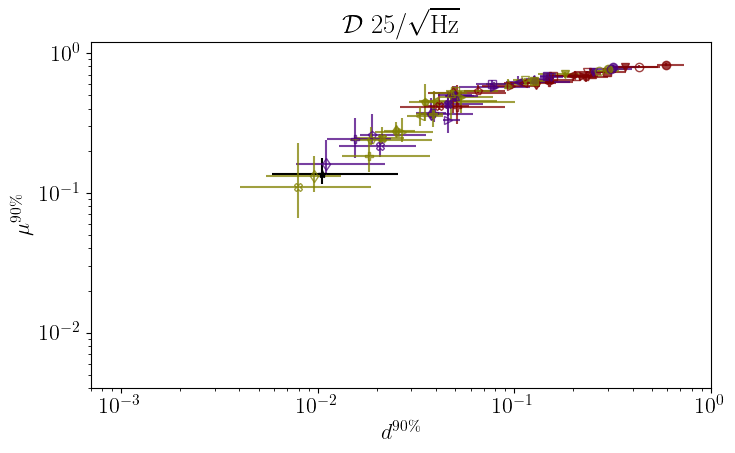}
    \includegraphics[width=\columnwidth]{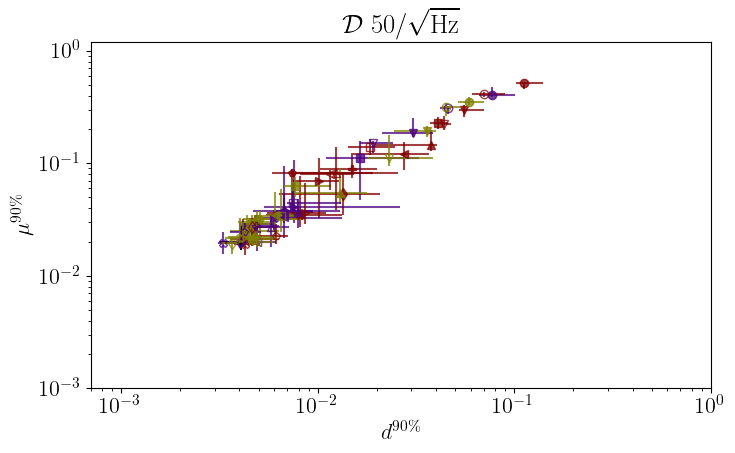}
    \includegraphics[width=\columnwidth]{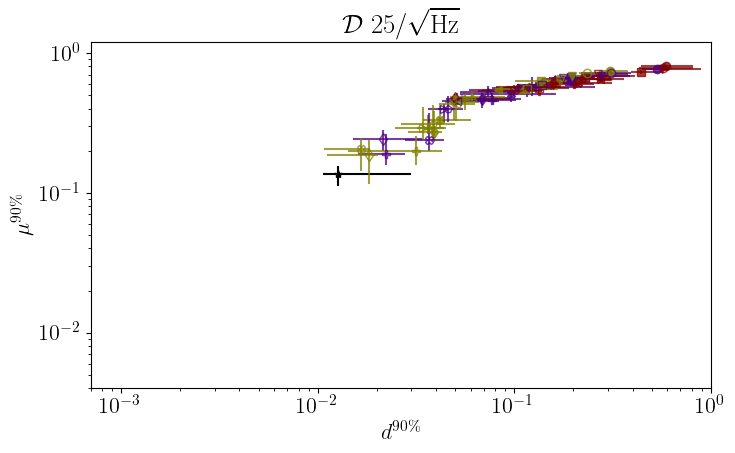}
    \caption{
        $90$th percentile of the $\mu$ distribution as a function of the same percentile of the $d$ distribution for the results
        presented in Fig.~\ref{fig:mismatch_vs_cost}. The distance is computed using $\Tsft = 1800,\,1024$~s 
        for the isolated (left column) and binary (right column) cases, respectively. Upper panels are computed using
        Gaussian data; lower panels are computed using real O3 Advanced LIGO data.
        Uncertainties are reported at the  68\% confidence level using bootstrapping.
        We only show the results for the higher depth values for simplicity. 
    }
    \label{fig:dist_vs_cost_real_data}
\end{figure*}

Additionally, we compute the distance between the injection point in Fig.~\ref{fig:dist_vs_cost_real_data}
for diagnostic purposes. The resulting values are safely below $1/\Tsft$.

Our results show that isolated signals can be followed up using configurations involving
about an order of magnitude less $\F$-statistic evaluations than previously realized,
both in Gaussian and real data. 
This amounts to about $10^{4}$ \mbox{$\F$-statistic} evaluations at 
$\depth = 50/\sqrt{\mathrm{Hz}}$ versus the \mbox{$1.8 \times 10^{5}$}
evaluations used by the standard configuration. For $\depth = 30 / \sqrt{\mathrm{Hz}}$,
the number of \mbox{$\F$-statistic} evaluations can be reduced down to
about $2 \times 10^{3}$.

A similar reduction is observed for the case of unknown sources in binary systems.
Here, however, the sensitivity depth is 20\% to 50\% lower; a smaller reduction
is expected for weaker signals. We also note that the initial prior shifting highly affects the reported performances. 
The expected deviation of a candidate from the 
parameters of a true signal is highly dependent on the characteristics of the pipeline 
producing the candidates in the first place. In this regard, our results are a
conservative estimate of  the achievable computing cost reduction,
and we expect them to be applicable to general all-sky searches.
These results are consistent with those reported in~\cite{Covas:2024pam}, where successful
follow-ups incur in a growing computing cost as signals become weaker and parameter-space regions
become broader.

\subsection{multistage follow-up\label{subsect:multi-stage}}

\begin{table}
    \centering
    \renewcommand*{\arraystretch}{1.5}
    \begin{tabular}{cccccc}
        \hline
         & $\mathcal{N}_{*}$ & $\Tcoh^{(0)}$ & $\Tcoh^{(1)}$ & $\Tcoh^{(2)}$ & $\Tcoh^{(3)}$\\
        \hline
        Isolated & $10^4$ & 0.5 & 5 & 50 & -- \\
        \hline
        Binary & $10^3$ & 0.5 & 1.6 & 5 & 15\\
        \hline
    \end{tabular}
    \caption{
        Coherence times in days for the follow-up of isolated and binary signals 
        calculated with Eq.~\eqref{eq:ladder_approx}. 
    }
    \label{tab:Tcoh_ladder}
\end{table}

We now test the performance of applying a coherence-time ladder
to the results of Sec.~\ref{subsect:stage-0}. To do so, we select 
\mbox{$\N_{*} = 10^{3}, 10^{4}$} for the binary and isolated
cases, respectively, and compute the coherence ladder starting from
$\Tcoh = 0.5~\mathrm{day}$ using Eq.~\eqref{eq:ladder_approx}.
The resulting ladders are listed in Table~\ref{tab:Tcoh_ladder}.
As discussed in the introduction, high coherence times may fail to recover
a CW signal due to unmodeled effects. To account for these, we limit the maximum coherence
time to 50 days in the isolated case to account for possible pulsar
glitches~\cite{10.1111/j.1365-2966.2011.18503.x,Ashton:2017wui} and to 
15 days in the binary case to account for
spin wandering~\cite{Whelan:2023,Mukherjee:2017qme,PhysRevD.105.022002}.

We compute $\muq$ and $d^{90\%}$ for the different stages as in Sec.~\ref{subsect:stage-0}.
Each stage takes as input the loudest candidate from the previous one.
Prior distributions for stage $j$ are built around the loudest candidate's
parameters following Table~\ref{tab:inj_pars_and_prior}, 
adjusting the priors' width according to $\Tcoh^{(j-1)}$. 

\begin{figure*}
    \centering
    \includegraphics[width=\columnwidth]{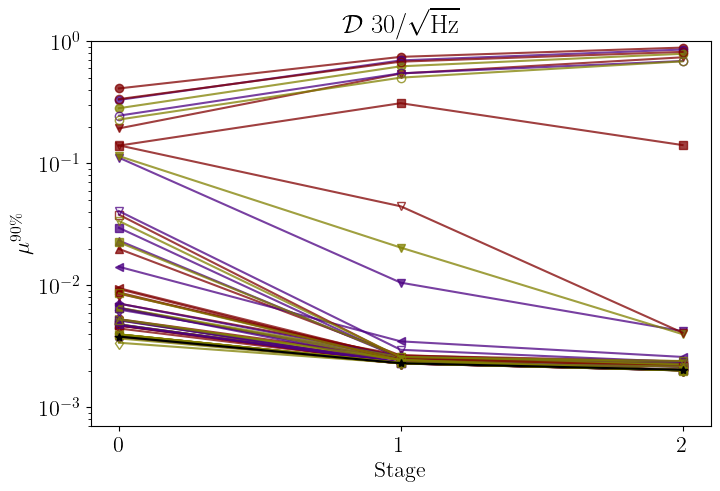}
    \includegraphics[width=\columnwidth]{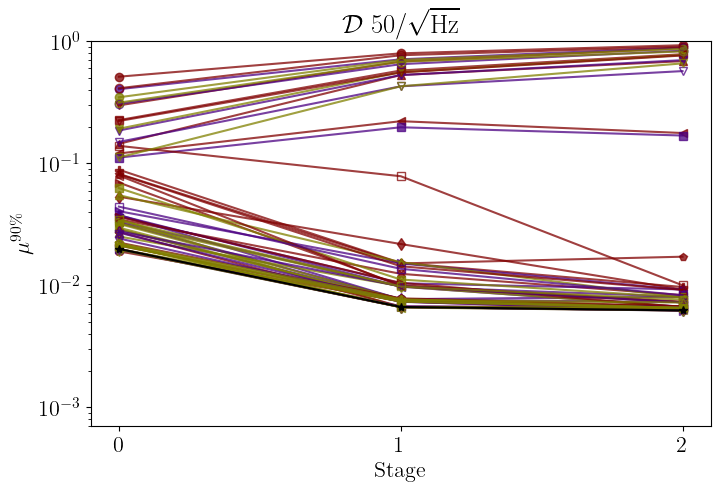}
    \includegraphics[width=\columnwidth]{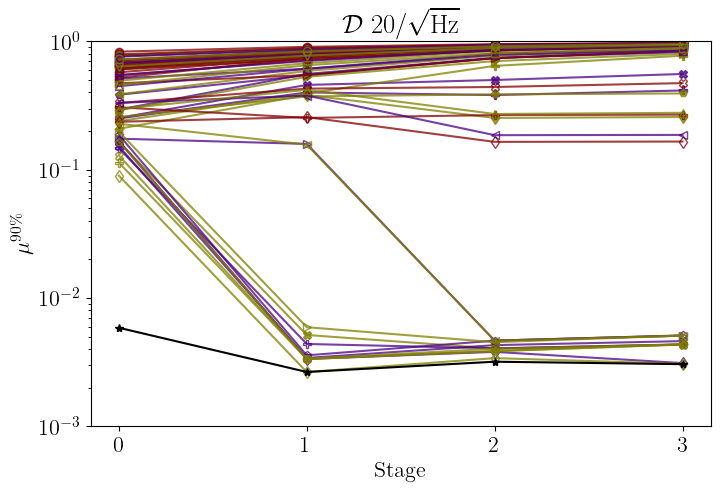}
    \includegraphics[width=\columnwidth]{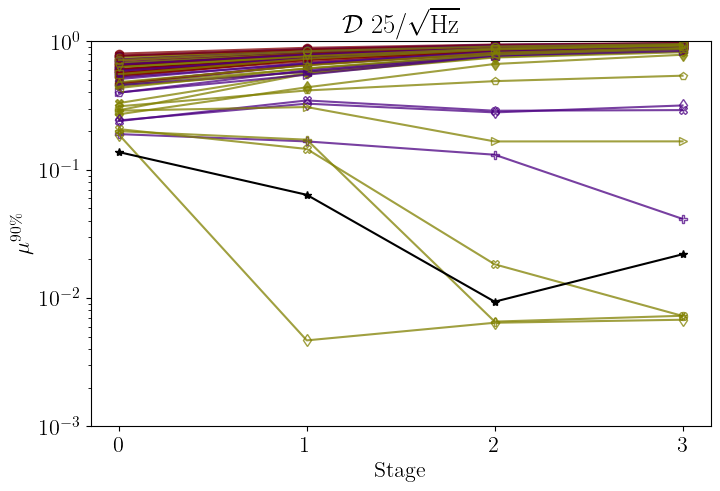}
    \caption{
        $90$th percentile  of the mismatch distribution after each stage of the
        coherence time ladder (see Table~\ref{tab:Tcoh_ladder}) on real Advanced LIGO data.
        The upper row corresponds to the isolated injection campaign.
        The lower row corresponds to the binary case. 
        Markers are consistent with
        the legend described in Fig.~\ref{fig:mismatch_vs_cost}.
    }
    \label{fig:multi-stage_real_data}
\end{figure*}

The evolution of $\muq$ and $d^{90\%}$ across follow-up stages in real data
is shown in Figs.~\ref{fig:multi-stage_real_data} and~\ref{fig:multi-stage_real_data_dist}.
The results for Gaussian data are shown in Appendix~\ref{app:multi_stage_fake_data}.
In this case, our analysis is limited to compare the behavior of different configurations 
to that of the standard configuration. 

Figure~\ref{fig:multi-stage_real_data} displays a clear trend on the effectiveness of 
MCMC configurations across the coherence-time ladder: most configurations achieving
$\muq \lesssim (0.1 - 0.2)$ in the first follow-up stage, regardless of the signal population,
are able to converge to the true signal parameters as the ladder progresses, as shown
in Fig.~\ref{fig:multi-stage_real_data_dist}. Note, however, that such a convergence
is conditioned on the steepness of the ladder.

\begin{figure*}
    \centering
    \includegraphics[width=\columnwidth]{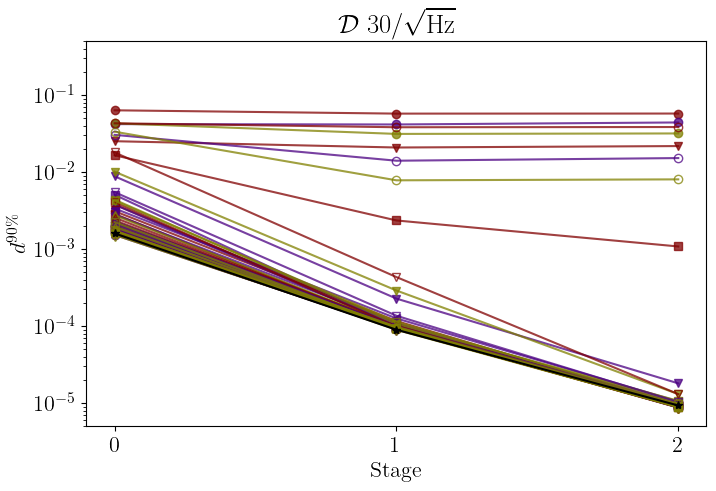}
    \includegraphics[width=\columnwidth]{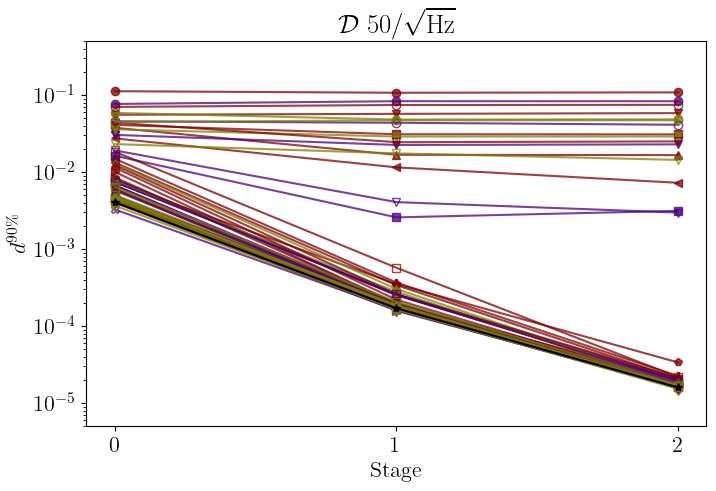}
    \includegraphics[width=\columnwidth]{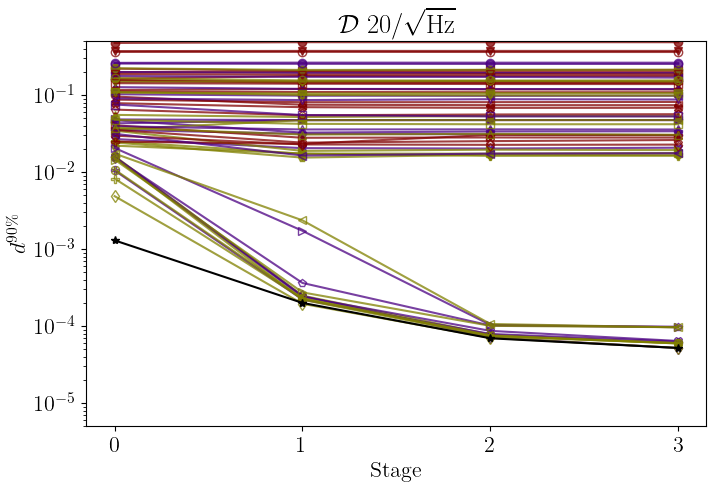}
    \includegraphics[width=\columnwidth]{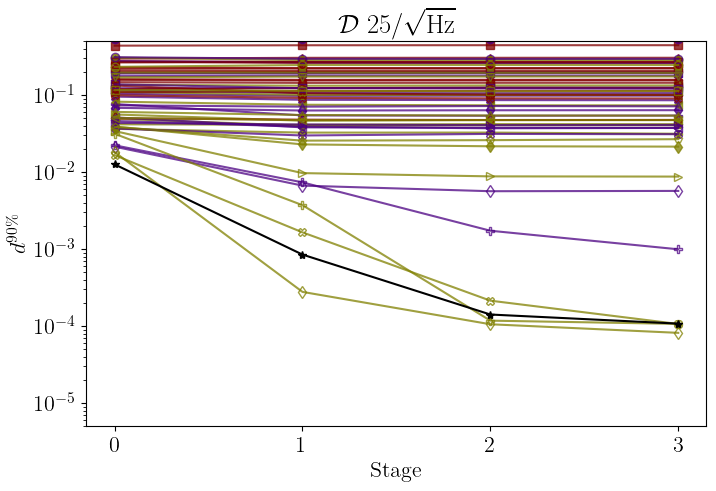}
    \caption{
        $90$th percentile  of the distance distribution after each stage of the
        coherence time ladder (see Table~\ref{tab:Tcoh_ladder}) on real Advanced LIGO data.
        The upper row corresponds to the isolated injection campaign.
        The lower row corresponds to the binary case.
        Markers are consistent with
        the legend described in Fig.~\ref{fig:mismatch_vs_cost}.
    }
    \label{fig:multi-stage_real_data_dist}
\end{figure*}

\subsection{Discussion\label{subsect:discussion}}

The results presented in Secs.~\ref{subsect:stage-0} and~\ref{subsect:multi-stage}
demonstrate the benefits of optimizing the configuration of the follow-up.
In previous works~\cite{KAGRA:2021una,PhysRevD.103.064017,PhysRevD.105.022002,
KAGRA:2022osp,Whelan:2023,LIGOScientific:2022enz,PhysRevD.106.102008},
a standard setup was used after verifying proper recovery of the target signal population.
Here, we have proved that a successful outcome, with comparable recovery performances,
can be obtained at a much lower computing cost.

We identify the configuration  \mbox{$\Nwalk=50,\Ntemp=1, \Ntot=300$}, 
corresponding to the black filled right triangle, as a good general default choice for the first follow-up stage, 
as it reduces the computing cost by a factor 12 and returns an  acceptable mismatch level 
at the probed sensitivity depths for isolated and binary signals in both Gaussian and real 
data. Cheaper configurations could be selected if one were interested in signals with
higher amplitudes (\mbox{$\depth \lesssim 30 / \sqrt{\mathrm{Hz}}$}). These results are
broadly consistent with those discussed in~\cite{Covas:2024pam} for a different
follow-up application.

Figure~\ref{fig:multi-stage_real_data} shows a successful application of our newly
proposed coherence time ladder, which is simpler to calculate and tends to return a slightly
lower number of steps than the previous proposal. We note, however, that the maximum
allowed  mismatch for a configuration to be effective across multiple ladder stages is smaller 
than for a single-stage follow-up. This is especially true for sources in binary systems:
due to their higher parameter-space dimensionality, only setups two to five times cheaper
than the standard configuration are able to follow the proposed ladder. A more favorable
ladder to specific follow-up applications can be constructed by properly tuning the selected
value of $\N_{*}$.

Taken together, these results allow for a variety of computationally efficient strategies.
For the case of single-stage follow-ups, the newly proposed configuration will be able to discard
noise outliers with the desired false-alarm probability. Upon completion of this first step,
slightly more expensive setups (by about a factor 2) can be used to further narrow down the
parameters of the candidate before starting the coherence time ladder. Under this scheme,
the cost of running a complete ladder on a candidate involves 3 to 5 stages, each of them using 
$10^{3}$ to, at most, $10^{5}$ $\F$-statistic evaluations.

These results were obtained at a follow-up false-dismissal probability of $\pfd = 10^{-5}$ at 
an amplitude corresponding to the 95\% detection probability of a search. In general, however, searches could
detect signals at higher sensitivity depths (lower amplitudes), for which the results here presented may be
not entirely applicable. We recommend the reader to take the results here presented as well-informed starting point,
and to calibrate specific setups for production use by means of a minimal injection campaign to account
for any unexpected behaviors of the follow-up caused by the dataset at hand.

\section{Impact on all-sky search sensitivities\label{sect:impact}}

We devote this section to justifying the importance of postprocessing and follow-up strategies
when it comes to improving the sensitivity of broad parameter-space searches for CWs. To do so,
we run an all-sky search on 40 \SI{0.1}{\hertz} representative frequency bands of Advanced LIGO
O3 data using the~\texttt{SkyHough} pipeline~\cite{Krishnan:2004sv, lalsuite} with the same
configuration as in~\cite{PhysRevD.106.102008}.  These 40 representative frequency bands, namely 
[66.325, 70.300, 71.275, 73.250, 75.350, 93.200, 99.375, 105.475,
109.600, 119.000, 130.475, 135.650, 142.100, 146.725, 158.225, 168.875,
173.075, 176.150, 181.000, 193.575, 201.325, 215.900, 238.475, 242.875,
247.850, 249.100, 260.200, 262.875, 269.500, 273.650, 280.000, 282.050,
301.000, 304.525, 309.550, 312.900, 322.950, 334.100, 334.900, 336.000] \unit{\hertz}, 
correspond to a subset of the bands used to estimate the sensitivity of the search.

In each band, an all-sky template bank covering a linear spindown range between 
\mbox{$\SI{-1e-9}{\hertz/\second}$} and \mbox{$\SI{1e-12}{\hertz/\second}$}
is analyzed  using different detection statistics as described in Sec.~IV~B~1
of~\cite{PhysRevD.106.102008}. The result per band is a toplist containing the
top $10^5$ candidates, which then are clustered using the algorithm presented
in~\cite{PhysRevD.103.064053}.

\begin{figure}
    \includegraphics[width=\columnwidth]{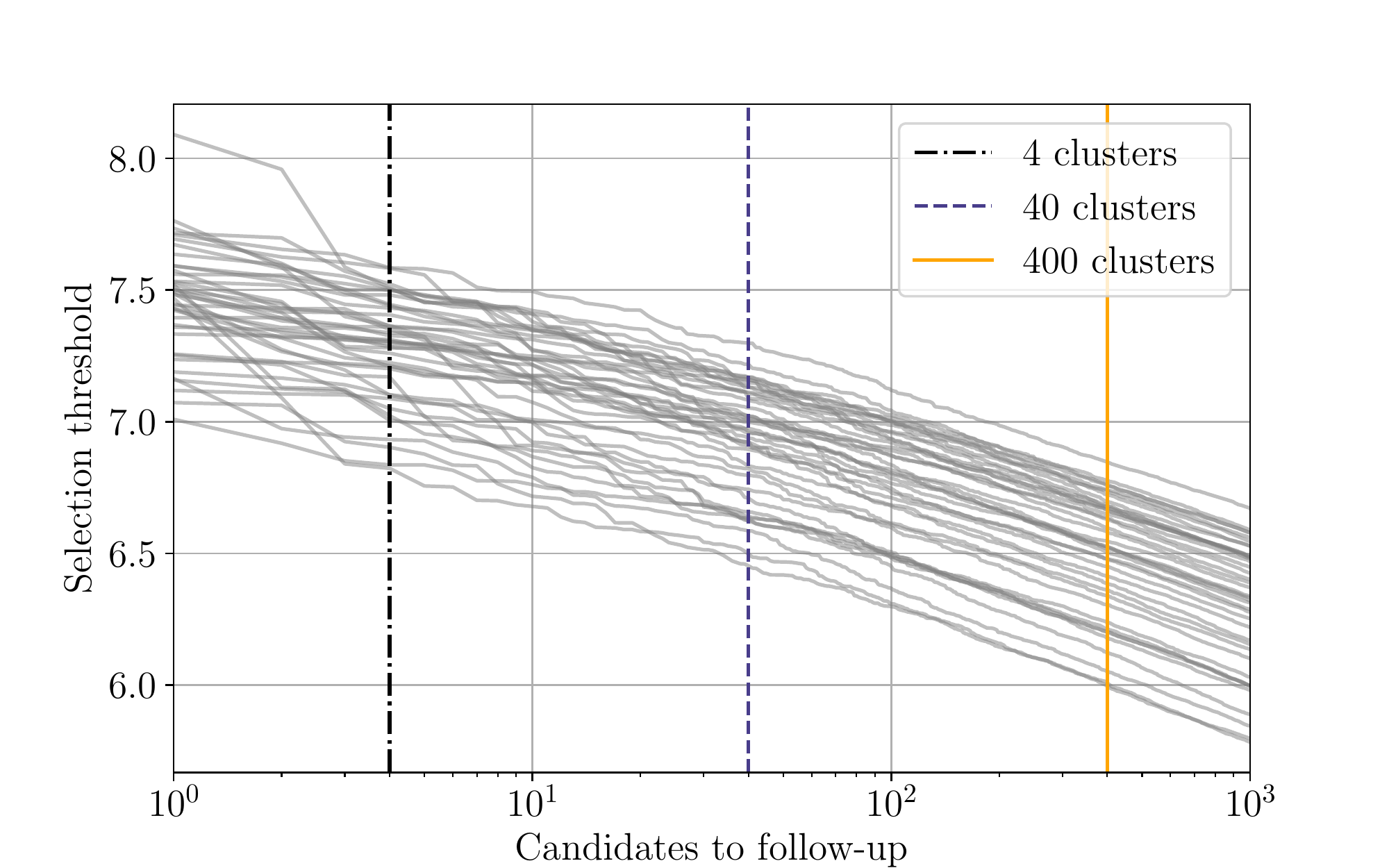}
    \caption{
        Statistical threshold corresponding to the last selected cluster for
        all the analyzed frequency bands as a function of the number of selected candidates.
        Each gray line represents a different frequency band.
        The quantity on the vertical axis corresponds to the critical ratio of the weighed normalized
        power, as described in~\cite{PhysRevD.106.102008}.
    }
    \label{fig:uls_thresholds}
\end{figure}

At this point, the search in~\cite{PhysRevD.106.102008} selected the top 40 clusters
in each band and estimated the search sensitivity by means of an injection campaign.
The number of selected clusters was significantly higher than previous searches
(cf. 1 cluster in O1 isolated~\cite{PhysRevD.96.062002},
1 cluster in O2 isolated~\cite{PhysRevD.100.024004}, 3 clusters in O2 all-sky
binary~\cite{PhysRevLett.124.191102}). In turn, the resulting sensitivity depth
increased by about 20\% with respect to said searches. Part of this sensitivity
increase is due to the duration of the dataset and the improved noise background;
part is due to the increase in the number of selected candidates as shown in
Fig.~\ref{fig:uls_thresholds}: An increased number of candidates corresponds to increasing
the false-alarm probability of the search, which in turn tends to increase the detected
fraction of a given population. This approach, however, runs into diminishing returns as 
the threshold falls off rather slowly with the number of candidates.

\begin{figure}
    \includegraphics[width=\columnwidth]{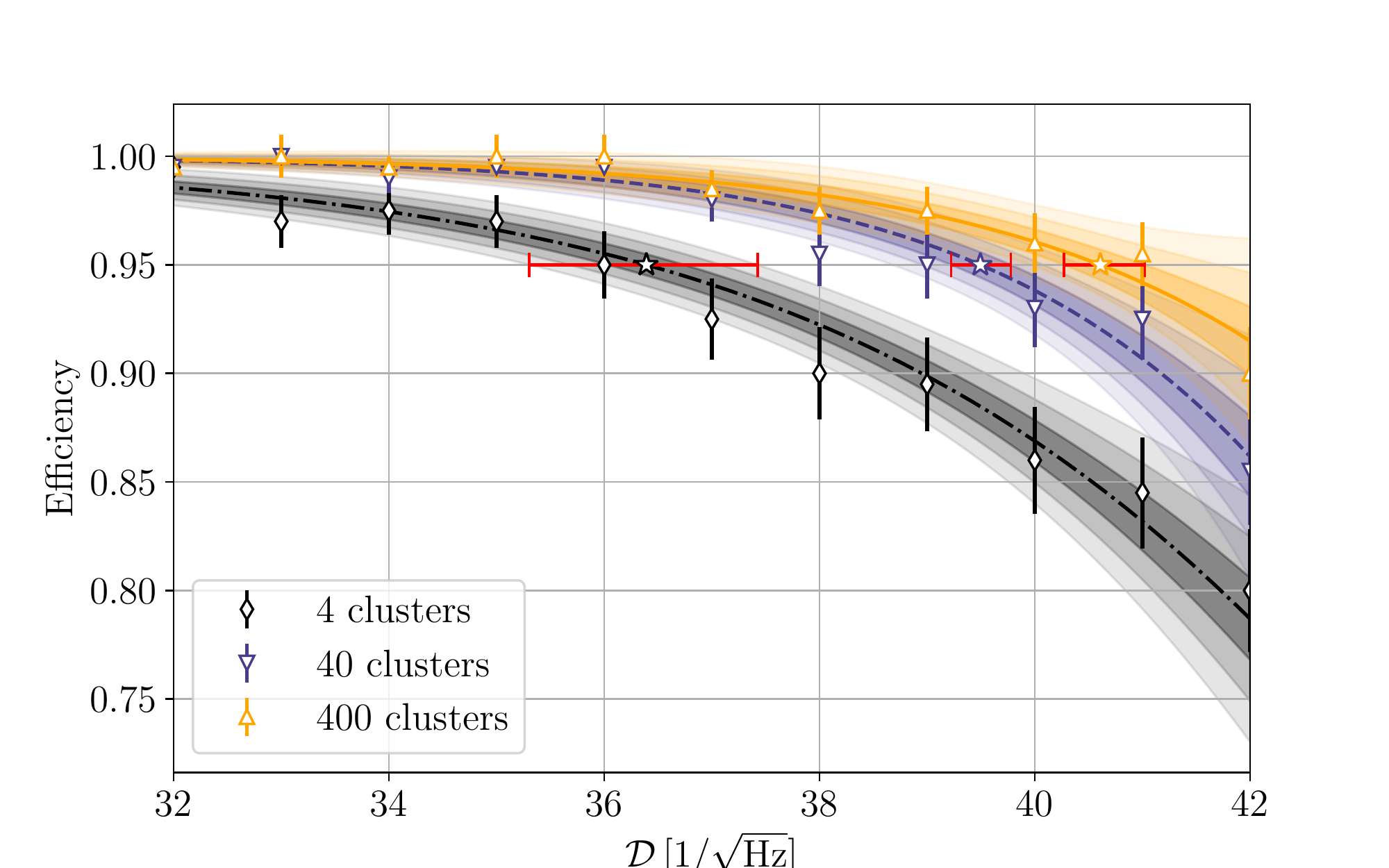}
    \caption{
        Example of sensitivity estimation fit. Each set of markers corresponds to a different
        threshold given by the last selected cluster. The solid, dashed, and dot-dashed lines represent
        the sigmoid fit. Shaded regions represent the 1, 2, and 3 sigma uncertainties as computed
        from the covariance matrix. We use one sigma for the 40 and 400 cases, and three sigma 
        for 4 as noise fluctuations make the fits very unstable.
        }
    \label{figs:uls_fit}
\end{figure}

We now evaluate the sensitivity impact of being able to follow up a higher number of CW candidates.
To do so, we simulate 200 CW signals at sensitivity depths ranging from 
$\mathcal{D} = 32/\sqrt{\mathrm{Hz}}$ to  $\mathcal{D} = 47/\sqrt{\mathrm{Hz}}$ 
and interpolate the sensitivity depth $\mathcal{D}^{95\%}$ corresponding to a $95\%$ detection probability
as explained in Sec.~V~B~2 of~\cite{PhysRevD.106.102008}. Uncertainties on $\mathcal{D}^{95\%}$ are 
computed using the covariance matrix of the fit. We follow a consistent detection criterion:
a simulated signal is considered detected if (1) the detection statistic of at least one of the 
selected clusters is above that of the last selected cluster in the all-sky search and (2) the loudest
candidate within said cluster is less than two parameter-space bins away from the injection point.
An example of this fitting procedure is shown in Fig.~\ref{figs:uls_fit}.

\begin{figure}
    \includegraphics[width=\columnwidth]{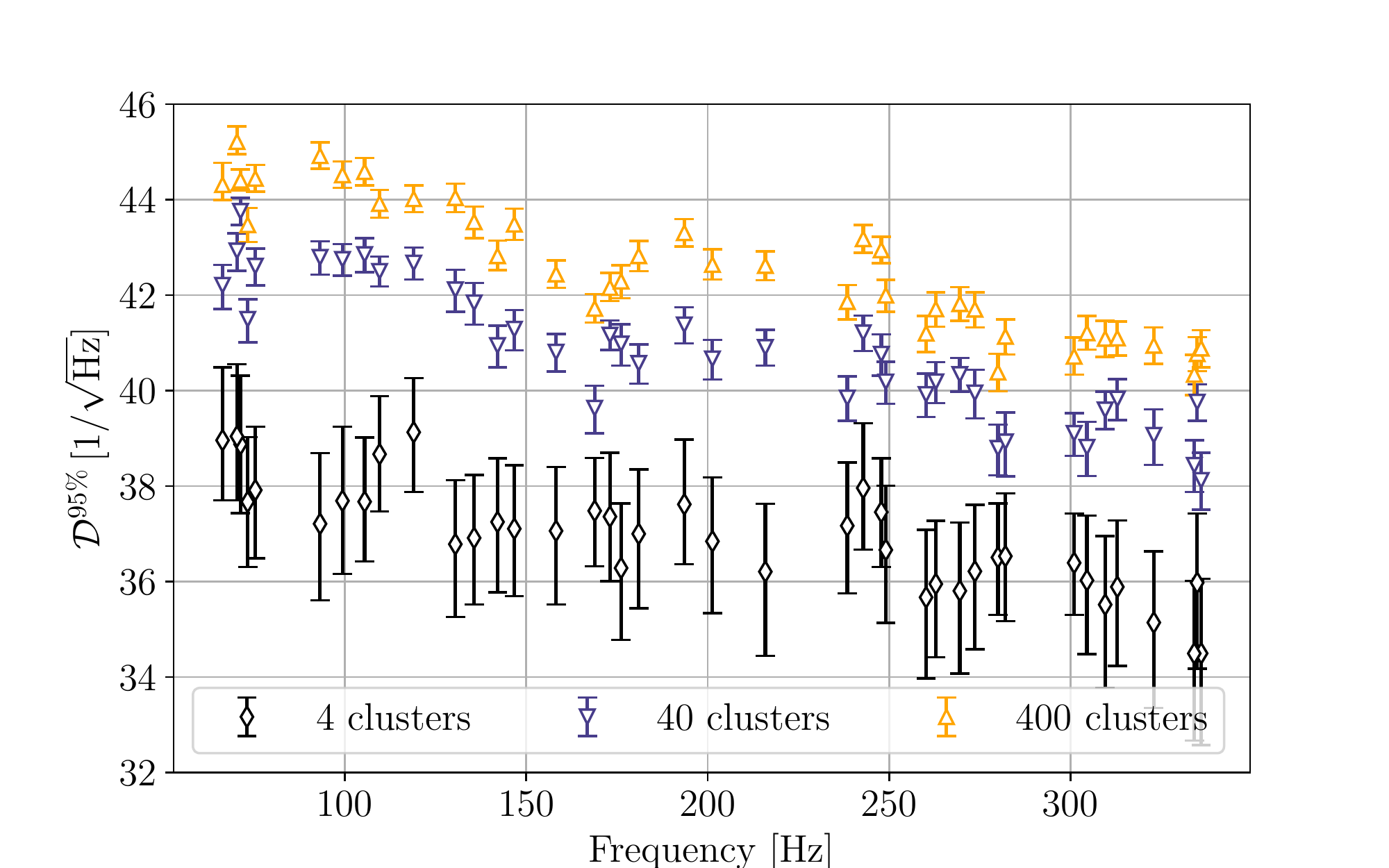}
    \caption{
        Sensitivity depths corresponding to a 95\% detection efficiency at different
        frequency bands. Results consistent with the original O3 search~\cite{PhysRevD.106.102008}
        are shown as downward triangles. Error bars correspond to 1 sigma fit uncertainties for the cases
        of 40 and 400 clusters. Due to the instability of the fit, 3 sigma uncertainties were used for the
        case of 4 clusters.
    }
    \label{figs:uls_sensitivity}
\end{figure}

The resulting sensitivity estimates are summarized in Fig.~\ref{figs:uls_sensitivity}. 
Increasing the number of selected clusters from 4 to 40 produces a median sensitivity increase
of about $10\%$ across the whole frequency band, with a maximum (minimum) increase of 15\% (6\%).
Further increasing from 40 to 400 clusters produces a median increase of about $4\%$, with a maximum
(minimum) increase of 7\% (1\%).

These results demonstrate that, overall, affordable follow-up strategies imply a better sensitivity 
for all-sky searches. This improvement, however, is likely to saturate upon reaching the bulk of 
the noise distribution, as suggested by the threshold behavior in Fig.~\ref{fig:uls_thresholds}.

\section{Conclusion\label{sect:conclusions}}

Blind searches for CW sources are in a privileged position to claim the first detection of a CW signal.
This is propitiated by the use of semicoherent methods~\cite{Tenorio:2021wmz}, which allow to 
evaluate broad parameter-space regions under an affordable computing budget by reducing 
the strictness with which a signal template is compared to the data~\cite{PhysRevD.93.044058,
PhysRevD.97.103020, Dergachev:2019wqa, Tenorio:2021wmz,Chua:2022ssg}.
Incidentally, this makes them robust to the stochastic nature of CW sources
such as NS stars, which are known to be affected by glitches~\cite{Ashton:2017wui}
and spin-wandering~\cite{Mukherjee:2017qme}.

The basic search strategy is to use affordable methods to select interesting candidates
which then are further evaluated using more sensitive follow-up methods in a hierarchical scheme.
Search sensitivity is thus limited by the number of candidates to be followed up.

In this work, we presented a new framework to evaluate the effectiveness of a follow-up configuration.
We apply this framework to \texttt{pyfstat}~\cite{PhysRevD.97.103020,Keitel:2021xeq}, a state-of-the-art
MCMC follow-up  widely accepted by the CW community~\cite{KAGRA:2021una,PhysRevD.103.064017,PhysRevD.105.022002,
KAGRA:2022osp, Whelan:2023,LIGOScientific:2022enz,PhysRevD.106.102008}. Additionally, we overall
simplify the setup of multistage follow-up strategy to ease their applicability to generic CW searches.
We test our proposals using simulated Gaussian and real O3 Advanced LIGO data, 
and we consider all-sky populations of CW sources both isolated and in binary systems.

We demonstrated significant reduction of the computational cost in the follow-up of generic CW candidates at realistic sensitivity
depths on real data. If we focus on relatively strong signals\mbox{$(\depth \lesssim 30 / \sqrt{\mathrm{Hz}})$}, 
this reduction reaches two orders of magnitude. 
These findings are consistent with~\cite{Covas:2024pam}, which tackles the problem
of broad parameter-space follow-ups. With this, we estimated the sensitivity improvement on an all-sky
search by increasing the number of follow-up outliers, which may reach up to a 15\% increase in
sensitivity depth at a comparable computing cost.

These results will have a positive impact on the resulting sensitivity of all-sky searches in 
the forthcoming observing runs of the  LIGO-Virgo-KAGRA collaboration.
 
\section*{Acknowledgments}
We thank Alessandro De Falco, Pia Astone, Alicia M. Sintes, David Keitel
for encouraging and fruitful discussions.
We thank Rafel Jaume for computing support and Karl Wette, Luca D'Onofrio,
the CW working group of the LIGO-Virgo-KAGRA Collaboration, Pep Covas, Reinhard Prix, Jasper Martins, and the anonymous 
referee for comments on the manuscript.
L.M. thanks the GRAVITY group at the University of the Balearic Islands
for their hospitality during the development of this project.
R.T. is supported by 
ERC Starting Grant No.~945155--GWmining, 
Cariplo Foundation Grant No.~2021-0555, 
MUR PRIN Grant No.~2022-Z9X4XS, 
MUR Grant ``Progetto Dipartimenti di Eccellenza 2023-2027'' (BiCoQ),
the ICSC National Research Centre funded by NextGenerationEU,
the Universitat de les Illes Balears (UIB);
the Spanish Agencia Estatal de Investigaci\'on grants PID2022-138626NB-I00,
RED2022-134204-E, RED2022-134411-T,
funded by MICIU/AEI/10.13039/501100011033, and ERDF/UE;
the MICIU with funding from the European Union NextGenerationEU/PRTR (PRTR-C17.I1);
the Comunitat Autonòma de les Illes Balears through the Direcci\'o General de Recerca, Innovaci\'o I
Transformaci\'o Digital with funds from the Tourist Stay Tax Law (PDR2020/11 - ITS2017-006), 
the Conselleria d'Economia, Hisenda i Innovaci\'o Grant No.~SINCO 2022/18146 (HiTech-IAC3-BIO)
cofinanced by the European Union and FEDER Operational Program 2021-2027 of the Balearic Islands.
The authors are grateful for computational resources
provided by LIGO Laboratory, supported by National Science Foundation Grants No.~PHY-0757058 and No.~PHY-0823459.
This research has made use of data or software obtained from the Gravitational Wave Open Science Center (gwosc.org),
a service of the LIGO Scientific Collaboration, the Virgo Collaboration, and KAGRA.
This material is based upon work supported by NSF's LIGO Laboratory which is a major facility fully funded by the
National Science Foundation, as well as the Science and Technology Facilities Council (STFC) of the United Kingdom,
the Max-Planck-Society (MPS), and the State of Niedersachsen/Germany for support of the construction of Advanced LIGO
and construction and operation of the GEO600 detector. Additional support for Advanced LIGO was provided by
the Australian Research Council. Virgo is funded, through the European Gravitational Observatory (EGO),
by the French Centre National de Recherche Scientifique (CNRS), the Italian Istituto Nazionale di Fisica Nucleare (INFN)
and the Dutch Nikhef, with contributions by institutions from Belgium, Germany, Greece, Hungary, Ireland, Japan, Monaco,
Poland, Portugal, Spain. KAGRA is supported by Ministry of Education, Culture, Sports, Science and Technology (MEXT),
Japan Society for the Promotion of Science (JSPS) in Japan; National Research Foundation (NRF) and Ministry of Science
and ICT (MSIT) in Korea; Academia Sinica (AS) and National Science and Technology Council (NSTC) in Taiwan.
This document has been assigned document number LIGO-P2400221.\\

\appendix
\section{MCMC NOISE DISTRIBUTION\label{app:noise_distribution}}
We empirically estimate the noise distribution of different MCMC follow-up configurations.
To do so, we run a follow-up using a subset of the  MCMC configurations in 
Table~\ref{tab:MCMC_setups} and priors from Table~\ref{tab:inj_pars_and_prior} (with width $\Tcoh=\Tsft$)
on 150  Gaussian-noise realizations. We select durations of \mbox{$\Tobs = 0.5, 1~\mathrm{year}$} in
order to consider both  the binary and isolated cases. The number of segments is 365 and 730,
respectively, in order to have \mbox{$\Tcoh = \SI{0.5}{\day}$}. 

\begin{figure}
    \centering
    \includegraphics[width=\columnwidth]{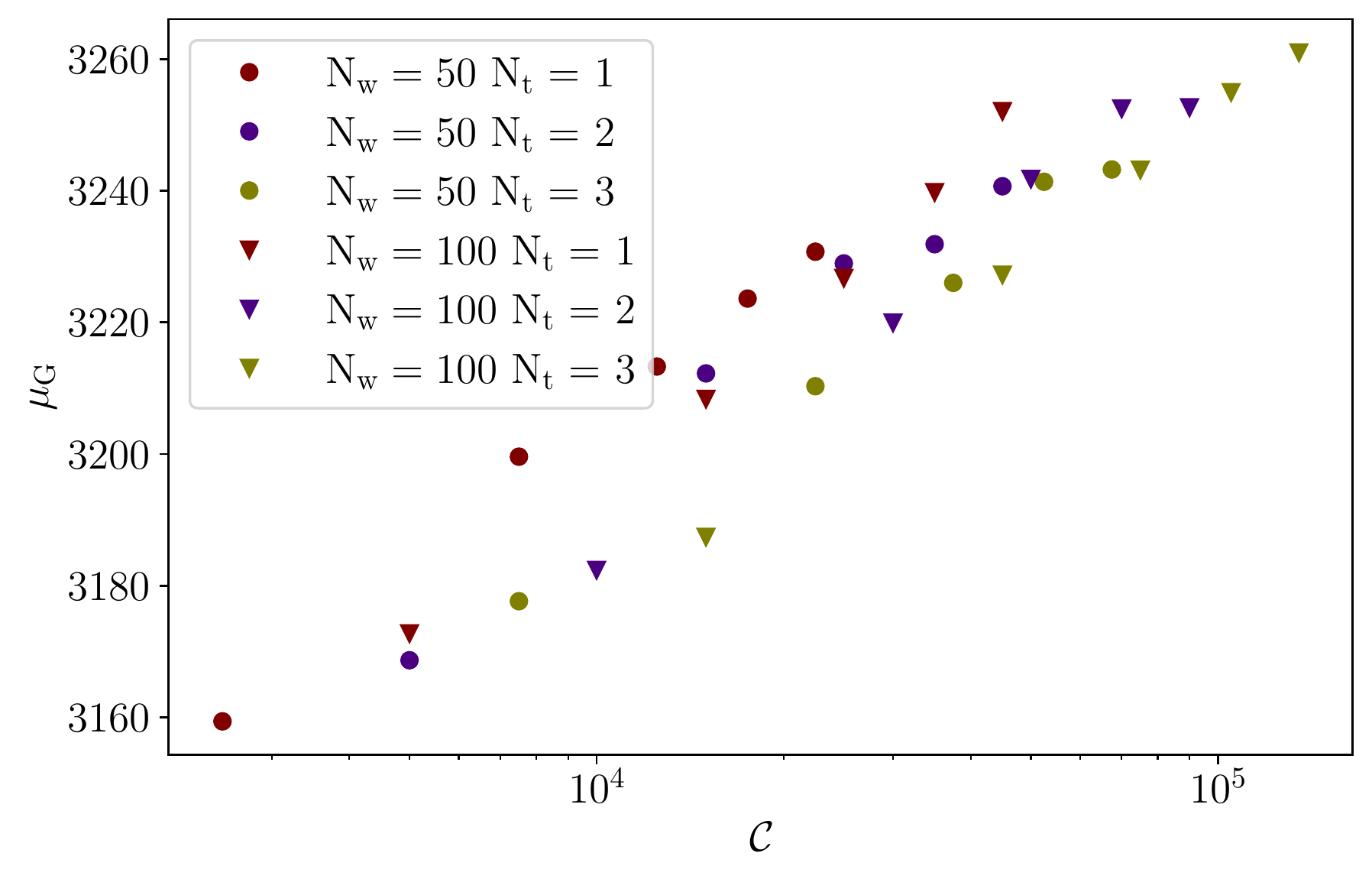}
    \includegraphics[width=\columnwidth]{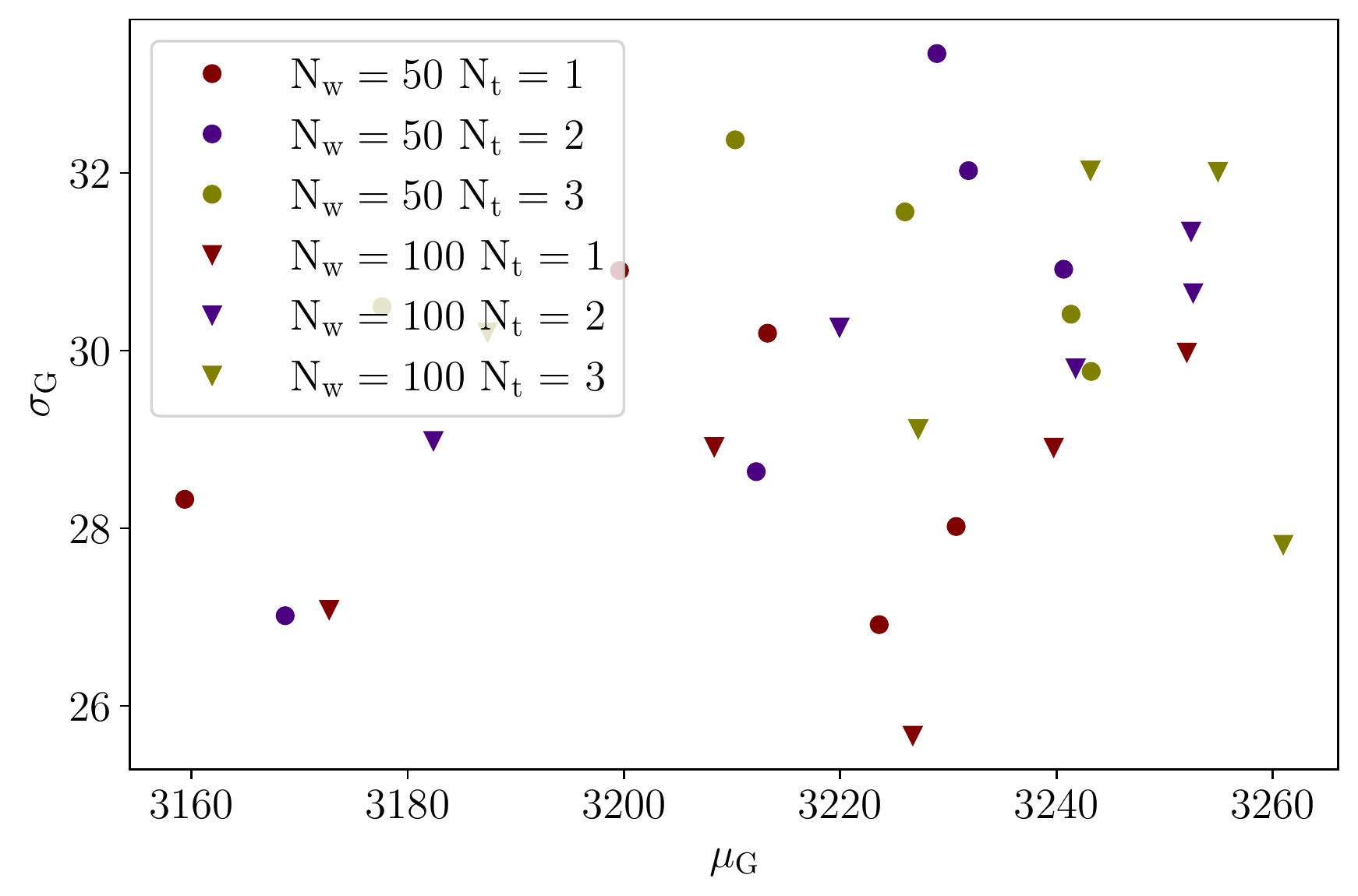}
    \caption{
        Upper panel: location parameter of the Gumbel distribution as a function of computing cost.
        Lower panel: scatter plot of location and scale parameters of a Gumbel distribution.
        Each pair of parameters is estimated by retrieving the loudest $\F$-statistic of 150
        MCMC runs on Gaussian noise and fitting a Gumbel distribution.
    }
    \label{fig:gumbel_fits}
\end{figure}

For each MCMC configuration, we retrieve $2\F_{\mathrm{max}}$ from each of the 150 noise 
realizations and fit a Gumbel distribution [Eq.~\eqref{eq:Gumbel}] as explained
in Sec.~\ref{sect:follow_up_introduction}. The resulting location $\mu_{\mathrm{G}}$ and
scale $\sigma_{\mathrm{G}}$ parameters for \mbox{$\Tobs = 1~\mathrm{year}$} are shown in 
Fig.~\ref{fig:gumbel_fits}. Similar results are observed for
\mbox{$\Tobs = 0.5~\mathrm{year}$}.

We find a strong correlation between the location parameter $\mu_{\mathrm{G}}$ and
(the logarithm) of the cost of the configuration. This is expected, as, in this case,
$\mu_{\mathrm{G}}$ grows with the logarithm of the size of the 
template bank~\cite{distromax}, which is proportional to the cost of a configuration.
The scale parameter fluctuates by about 10\% around an average value of $\sim 30$.

The noise distribution of the $\F$-statistic in Gaussian noise depends only
on $\Nseg$ regardless of the number of detectors, their PSD, and the observing
time~\cite{PhysRevD.58.063001,PhysRevD.72.063006,Prix:2009tq}. In particular, the results
generated in this appendix can be used for both the isolated and binary injection campaigns shown in
Figs.~\ref{fig:mismatch_vs_cost}~and~\ref{fig:mismatch_vs_cost_real_data}.
We note, however, that the behavior of outliers on real data is difficult to predict as
it strongly depends on their specific cause.\\

\section{MULTISTAGE FOLLOW-UP WITH SIMULATED DATA\label{app:multi_stage_fake_data}}
\begin{figure*}
    \centering
    \includegraphics[width=\columnwidth]{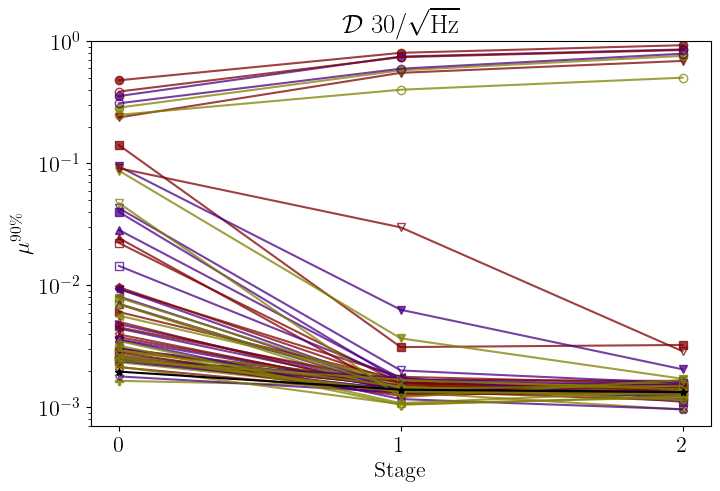}
    \includegraphics[width=\columnwidth]{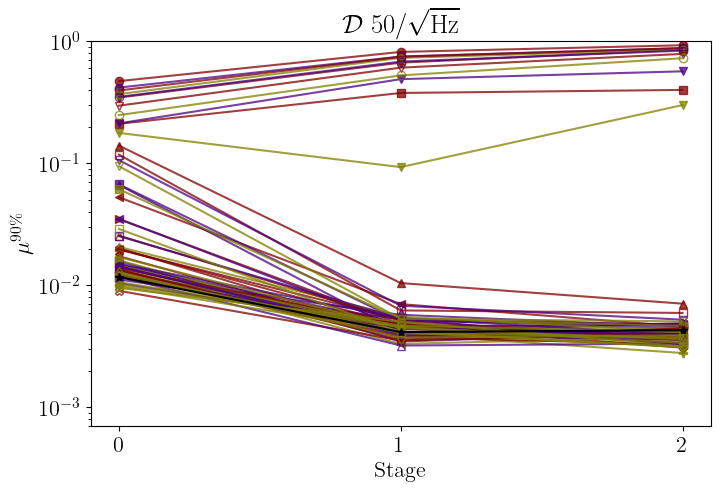}
    \includegraphics[width=\columnwidth]{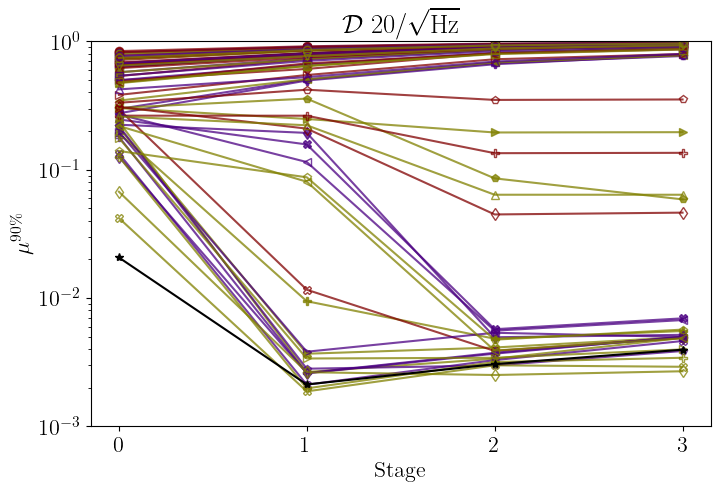}
    \includegraphics[width=\columnwidth]{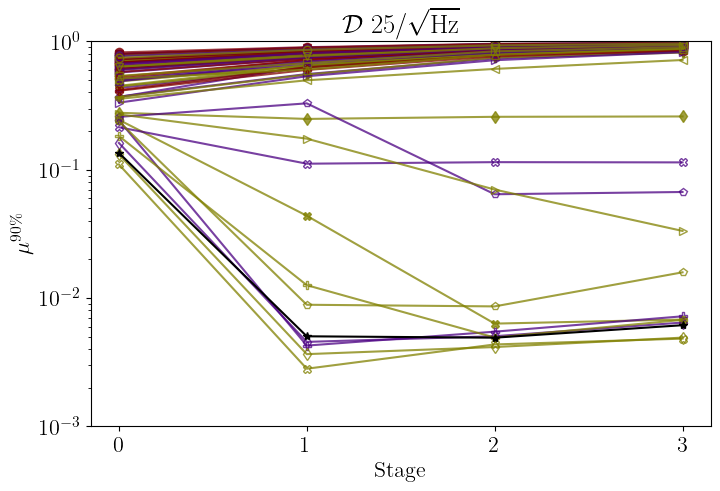}
    \caption{
        Same caption as Fig.~\ref{fig:multi-stage_real_data}, but using Gaussian data.
    }
    \label{fig:multi-stage_fake_data}
\end{figure*}

\begin{figure*}
    \centering
    \includegraphics[width=\columnwidth]{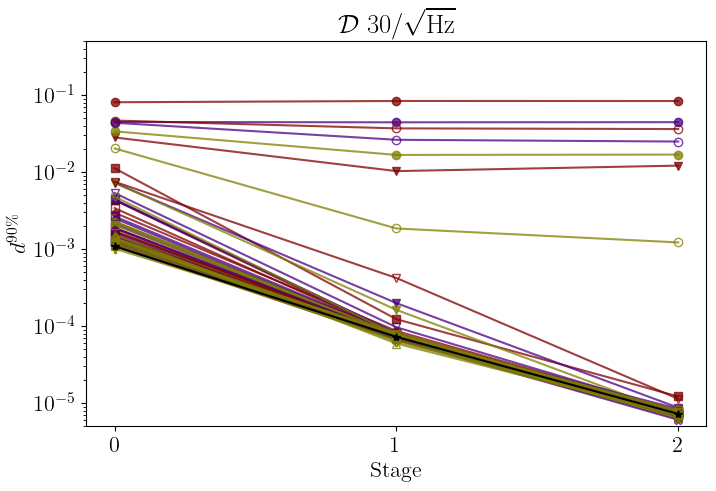}
    \includegraphics[width=\columnwidth]{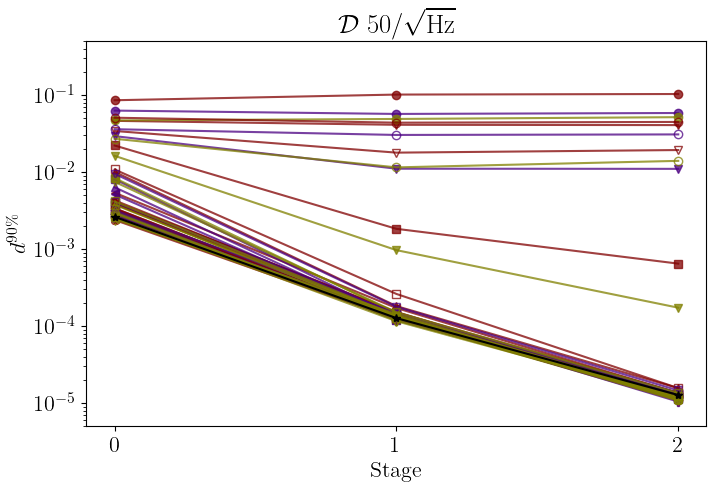}
    \includegraphics[width=\columnwidth]{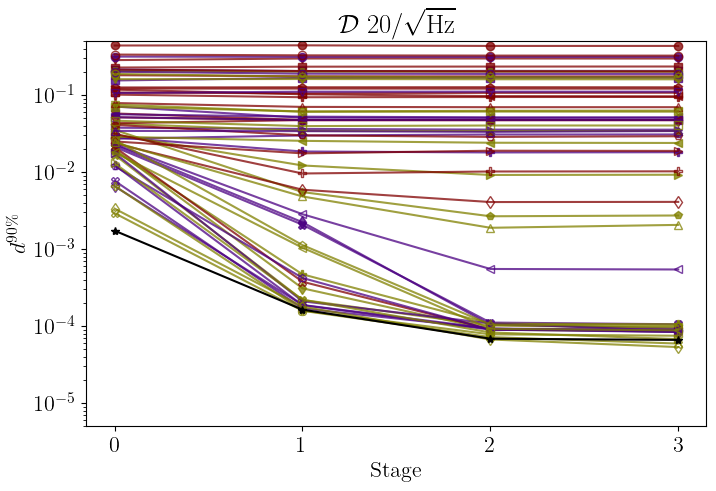}
    \includegraphics[width=\columnwidth]{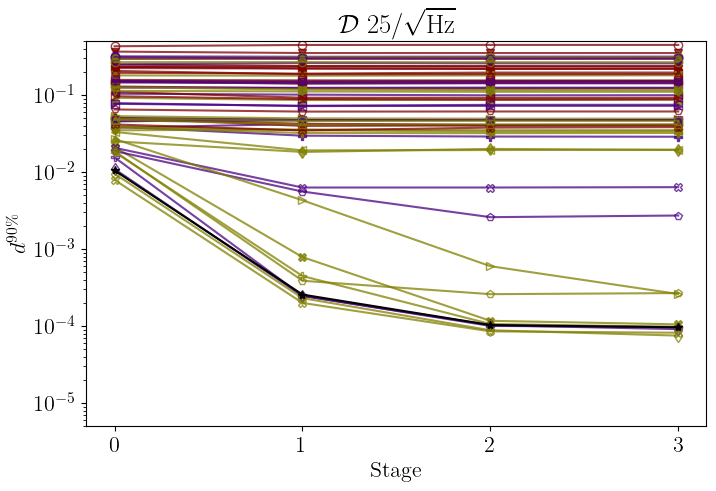}
    \caption{
        Same caption as Fig.~\ref{fig:multi-stage_real_data_dist}, but using Gaussian data.
    }
    \label{fig:multi-stage_fake_data_dist}
\end{figure*}

As mentioned in Sec.~\ref{sect:optimizing_setup}, we show in Fig.~\ref{fig:multi-stage_fake_data}
and Fig.~\ref{fig:multi-stage_fake_data_dist} the mismatch and distance results for a multistage
follow-up on simulated Gaussian data.

\bibliography{references}

\end{document}